\documentclass[aps,prb,twocolumn,amsmath,%
showpacs,showkeys,floatfix]{revtex4}

\usepackage{epsfig}
\usepackage[english]{babel}
\usepackage{graphicx}
\usepackage{graphics}
\usepackage{amsmath}
\usepackage{url}
\usepackage{amssymb}

\begin{document}

\renewcommand{\figurename}{FIG.}
\newcommand{\vc}[1]{\mathbf{#1}}

\title{Modeling of the electronic state of the High-Temperature
Superconductor LaCuO: Phonon dynamics and charge response}
\author{Claus Falter}
\email[Email to: ]{falter@nwz.uni-muenster.de}
\author{Thomas Bauer}
\author{Frank Schnetg\"oke}
\affiliation{Institut f\"ur Festk\"orpertheorie, Westf\"alische
Wilhelms-Universit\"at,\\ Wilhelm-Klemm-Str.~10, 48149 M\"unster,
Germany}
\date{\today}

\begin{abstract}
A modeling of the normal state of the p-doped high-temperature
superconductors (HTSC's) is presented. This is achieved
starting from a more conventional metallic
phase for optimal- and overdoping and passing via the
underdoped to the insulating state by consecutive orbital selective
compressibility-incompressibility transitions in terms of sum
rules for the charge response. The modeling is substantiated by
corresponding phonon calculations.  Extending investigations of
the full dispersion and in particular of the strongly doping
dependent anomalous phonon modes in LaCuO, which so far underpin
our treatment of the density response of the electrons in the
p-doped HTSC's, gives additional support for the modeling of the
electronic state, compares well with recent experimental data and
predicts the dispersion for the overdoped regime. Moreover, phonon
densities of states have been calculated and compared for the
insulating, underdoped, optimally doped and overdoped state of
LaCuO. From our modeling of the normal state a consistent picture
of the superconducting phase also can be extracted qualitatively
pointing in the underdoped regime to a phase ordering transition.
On the other hand, the modeling of the optimal and overdoped state
is consistent with a quasi-particle picture with a well defined
Fermi surface.  Thus, in the latter case a Fermi surface
instability with an evolution of pairs of well defined
quasiparticles is possible and can lead to a BCS-type ordering.
So, it is tempting to speculate that optimal $T_C$ in the HTSC's
marks a crossover region between these two forms of ordering.
\end{abstract}



\maketitle

\section{Introduction}\label{SecOne}
In our microscopic theoretical description of the charge response,
electron-phonon interaction (EPI) and lattice dynamics of the
high-temperature superconductors (HTSC's) in the framework of the
linear response approach we focus on the specific features of their
solid-state chemistry. This is different from the overwhelming number
of attempts relying on simplified models with strongly reduced model
Hamiltonians. Nevertheless, this seems necessary since the phenomena
observed in the cuprates hardly even occur in the other known
compounds. One important feature neglected for example in the popular
models of Hubbard type is the proximity of the HTSC's, even in the
optimally doped case, to the layered ionic crystals where besides
short-ranged correlations strong long-range Coulomb interaction and
strong (nonlocal) EPI exists. In the absence of full ab initio methods
for the HTSC's to tell us what the correct starting point would be,
such insight is essential, besides the proximity of the materials to
an antiferromagnetic state. Models with strongly reduced electronic
degrees of freedom may be useful for the discussion of theoretical
ideas in general rather than for detailed calculations of the material
specifica of the cuprates.

The strong ionic nature of the HTSC's is modeled in our approach
by an ab-initio rigid ion model (RIM), used as a reference frame
for the local ionic, rigid charge response and the EPI,
respectively. The important and characteristic nonlocal, non-rigid
contribution to the electronic density response and the EPI in the
HTSC's is calculated in terms of microscopically well defined
charge- and dipole fluctuations (CF's and DF's). The latter are
excitable on the
electronic shells of the ions in the crystal. This is, because
most of the charge
in the HTSC's contributing to bonding is concentrated on the
latter. Additionally, covalent metallic features of bonding are
approximatively taken into account \cite{Ref01,Ref02}. In the
calculations a sufficiently broad orbital freedom (Cu3d, Cu4s and
O2p), the full three-dimensional Coulomb interaction as well as
the short-ranged local repulsions, in particular the important
on-site repulsion mediated by the localized Cu3d orbitals, is
considered quantitatively. All the resulting couplings appearing
in the dynamical matrix and the EPI are microscopically well
defined and can be calculated. The results obtained so far agree
well with the experimental phonon dispersion. For a recent review,
see Ref. \onlinecite{Ref01}.

The model for the electronic polarizability matrix
\cite{Ref01,Ref03} $\Pi_{\kappa\kappa'}(\vc{q})$ of the
optimally-, and overdoped metallic state of the HTSC's, describing
the kinetic part of the charge response, is obtained in a first
step from the Kohn-Sham parameters (usually interpreted as
bandstructure) within the local density approximation (LDA) of
density functional theory (DFT). The indices $\kappa$, $\kappa'$
denote the orbital degrees of freedom in the elementary cell of
the crystal, i.e. Cu3d, Cu4s and O2p. $\vc{q}$ is a wavevector in
the first Brillouin zone. However, such a model, applied to LaCuO
overestimates the coupling along the $c$-axis and thus
underestimates the real anisotropy in the cuprates. Modifications
to improve the charge response along the ionic $c$-axis have been
investigated in Ref. \onlinecite{Ref03,Ref04}. Given the nearly
two-dimensional electronic structure of the cuprates and thus a
very weak interlayer coupling, electron dynamics and phonon
dynamics will be on the same time-scale in a small region around
the $c$-axis. This needs a non-adiabatic treatment with
dynamical screening of the bare Coulomb interaction and
phonon-plasmon mixing in the metallic state of the HTSC's
\cite{Ref01,Ref04}.

In the absence of a rigorous quantitative description for the
insulating (undoped) state of the HTSC's , the model for
$\Pi_{\kappa\kappa'}$ has been designed to be consistent with the
characteristic electronic charge response for an insulator with a gap
in the charge excitation spectrum. This is achieved by
fulfilling rigorous sum rules
for the density response in terms of the polarizability matrix in the
long-wavelength-limit \cite{Ref02,Ref05}. These sum rules, see Section
\ref{SecThree}, can be considered as orbital resolved closed forms to
represent the change of the charge response in a metal-insulator
transition in terms of the electronic polarizability or the
compressibility of the electronic system, respectively, as a primary
tool to characterize the ground state.

Finally, a suitable model is proposed for the
underdoped state of the p-doped HTSC's, where a
description of the low-energy excitations remains a theoretical
challenge and where the question how the metallic charge conduction
emerges from the doped insulator is poorly understood. This model
describes a novel metallic state with an
insulator-like, incompressible charge response related to the
localized Cu3d orbitals at the Cu sublattice interspersed with a
metallic, compressible charge response related to the more delocalized
O2p orbitals at the oxygen sublattices of the CuO plane. In this way a
partial ordering of the conducting carriers in the
CuO plane is obtained. With
regard to our modeling of the metallic state for the optimally and
overdoped HTSC's such an ergodicity breaking
delocalization-localization transition in the underdoped state in
terms of the compressibility means a compartmentalization of
configuration space. This means that some parts of configuration space
cannot be approached or are hardly accessible to the particles.
On the other hand, this should be accompanied by a reciprocal
compartmentalization of momentum space, according to the uncertainty
relationship $\Delta x \cdot \Delta k \sim 1$. Indeed, angle-resolved
photoemission spectroscopy (ARPES) studies in underdoped cuprates
\cite{Ref06,Ref07,Ref08}, reporting a ''two component'' electronic
structure with a pseudogap around $(\pi,0)$ and only part of the Fermi
surface surviving as an ''arc'' around the node, point to
compartmentalization of momentum space too.
The compartmentalization of momentum space counteracts the reduction
of the electron scattering rate brought about by the Pauli-principle
in a Fermi-liquid, at least partially. Thus, the scattering rate
and the life time of the carriers no longer will follow the well
known scaling with energy and temperature of a Fermi-liquid. Consequently,
the normal state should display non-Fermi-liquid properties
in our model for the underdoped state of the cuprates, also
seen in the experiment.
In the corresponding superconducting state an increase of the
kinetic energy as in BCS by the ''smearing'' of the Fermi surface
would not be present because the latter is not developed.
On the other hand, phase coherence
in a phase ordering scenario would enhance the mobility of
pairs and decrease the kinetic energy.
Such a description of a
delocalization-localization crossover of the electronic structure
(e.g. due to doping) accompanied by a pseudogap phenomenon in terms of
consecutive orbital-selective compressibility-incompressibility
transitions in the presence strong nonlocal EPI may also be useful to
model the charge correlations in other systems besides the cuprates
characterized by a mixture of localized and delocalized orbitals.

The article is organized as follows. In Section \ref{SecTwo} the
theory and modeling is briefly reviewed to set the frame. Section
\ref{SecThree} presents our calculated results of the phonon
dispersion for the underdoped and overdoped state of LaCuO. Moreover,
the phonon anomalies related to the high-frequency oxygen
bond-stretching modes and the phonon density of states for the
insulating, underdoped, optimally doped and overdoped state of LaCuO
are calculated. A discussion of the modeling of the electronic state
in the cuprates consistent with the corresponding phonon calculations
is provided. Finally, a summary will be presented in Section
\ref{SecFour}.

\section{Sketch of the Theory and modeling}\label{SecTwo}
From a general point of view our treatment of the electronic
density response and lattice dynamics in terms of DF's and CF's
can be considered as a microscopic (semi-ab-initio) implementation
of the phenomenological dipole-shell model or the charge-fluctuation
models, respectively. For a general formulation of phenomenological
models for lattice dynamics that use localized electronic variables
as adiabatic degrees of freedom, see for example
Ref. \onlinecite{Ref09}.
This formulation covers shell models, bond-charge
models and charge-fluctuation models. While in this approach
the coupling coefficients are treated as empirical fitting parameters,
the essential point in our microscopic scheme is that all the couplings
can be calculated.

In the following a survey of the theory and modeling is presented.
A detailed description can be found in Ref. \onlinecite{Ref05} and in
particular in Ref. \onlinecite{Ref10} where the calculation of the coupling
parameters is presented.

The local part of the electronic charge
response and the EPI is approximated in the spirit of the
quasi-ion approach \cite{Ref11} by an ab initio rigid ion model
(RIM) taking into account covalent ion softening in terms of
(static) effective ionic charges calculated from a tight-binding
analysis (TBA). In addition, scaling of the short-ranged part of
certain pair potentials between the ions is performed to simulate
further covalence effects in the calculation in such a way that
the energy-minimized structure is as close as possible to the
experimental one \cite{Ref12}. Structure optimization and energy
minimization is very important for a reliable calculation of the
phonon dynamics through the dynamical matrix.

The RIM with the corrections just mentioned then serves as an
unbiased reference system for the description of the HTSC's and
can be considered as a first approximation for the insulating
state of these compounds.  Starting with such an unprejudiced
rigid reference system non-rigid electronic polarization processes
are introduced in form of more or less localized electronic
charge-fluctuations (CF's) at the outer shells of the ions.
Especially in the metallic state of the HTSC's the latter dominate
the nonlocal contribution of the electronic density response and
the EPI and are particularly important in the CuO planes. In
addition, \textit{anisotropic} dipole-fluctuations (DF's) are
admitted in our approach \cite{Ref03,Ref10}, which prove to be
specificly of interest for the ions in the ionic layers mediating
the dielectric coupling and for the polar modes. Thus, the basic
variable of our model is the ionic density which is given in the
perturbed state by

\begin{equation}\label{Eq1}
\rho_\alpha(\vc{r},Q_\lambda, \vc{p}_\alpha) =
\rho_\alpha^0(r) + \sum_{\lambda}Q_\lambda \rho_\lambda^\text{CF}(r)
+ \vc{p}_\alpha \cdot
\hat{\vc{r}} \rho_\alpha^\text{D}(r).
\end{equation}

$\rho_\alpha^0$ is the density of the unperturbed ion, as used in the
RIM, localized at the sublattice $\alpha$ of the crystal and moving
rigidly with the latter under displacement. The $Q_\lambda$ and
$\rho^\text{CF}_\lambda$ describe the amplitudes and the form-factors
of the CF's and the last term in Eq. \eqref{Eq1} represents the dipolar
deformation of an ion $\alpha$ with amplitude (dipole moment)
$\vc{p}_\alpha$ and a radial density distribution
$\rho_\alpha^\text{D}$.  $\hat{\vc{r}}$ denotes the unit vector in the
direction of $\vc{r}$. The $\rho^\text{CF}_\lambda$ are approximated
by a spherical average of the orbital densities of the ionic shells
calculated in LDA taking self-interaction effects (SIC) into account.
The dipole density $\rho_\alpha^\text{D}$ is obtained from a modified
Sternheimer method in the framework of LDA-SIC \cite{Ref10}. All
SIC-calculations are performed for the average spherical shell in
the orbital-averaged form according to Perdew and Zunger \cite{Ref13}.
For the correlation part of the energy per electron $\varepsilon$ the
parametrization given in Ref. \onlinecite{Ref13} has been used.

The total energy of the crystal is
obtained by assuming that the density can be
approximated by a superposition of overlapping densities
$\rho_\alpha$. The $\rho_\alpha^0$ in Eq. \eqref{Eq1}
are also calculated
within LDA-SIC taking environment effects, via a Watson sphere
potential and the calculated static effective charges of the ions into
account. The Watson sphere method is only used for the oxygen ions
and the depth of the Watson sphere potential is set as the Madelung
potential at the corresponding site.
Such an approximation holds well in the HTSC's
\cite{Ref12,Ref14}. As a general rule, partial covalence reduces the
amplitude of the static effective charges in mixed ionic-covalent
compounds like the HTSC's, because the charge transfer from the
cations to the anions is not complete as in the entirely ionic
case. Finally, applying the pair-potential approximation we get for
the total energy:

\begin{equation}\label{Eq2}
E(R,\zeta) = \sum_{\vc{a},\alpha} E_\alpha^\vc{a}(\zeta)
+\frac{1}{2}\sum_{(\vc{a},\alpha)\neq(\vc{b},\beta)}
\Phi_{\alpha\beta}
\left(\vc{R}^\vc{b}_\beta-\vc{R}^\vc{a}_\alpha,\zeta\right).
\end{equation}

The energy $E$ depends on both the configuration of the ions $\{R\}$
and the electronic (charge) degrees of freedom (EDF) $\{\zeta\}$ of
the charge density, i.e. $\{Q_\lambda\}$ and $\{\vc{p}_\alpha\}$ in
Eq. \eqref{Eq1}.  $E_\alpha^\vc{a}$ are the energies of the single
ions. $\vc{a}$, $\vc{b}$ denote the elementary cells and $\alpha$,
$\beta$ the corresponding sublattices.  The second term in
Eq. \eqref{Eq2} is the interaction energy of the system, expressed in
terms of \textit{anisotropic} pair-interactions
$\Phi_{\alpha\beta}$. Both $E_\alpha^\vc{a}$ and $\Phi_{\alpha\beta}$
in general depend upon $\zeta$ via $\rho_\alpha$ in Eq. \eqref{Eq1}.

The pair potentials in Eq. \eqref{Eq2} can be seperated into long-ranged
Coulomb contributions and short-ranged terms as follows:

\begin{align}\nonumber
\Phi_{\alpha\beta}(\vc{R},\zeta) =& \frac{\mathcal{Z}_\alpha \mathcal{Z}_\beta}{R}
-(\mathcal{Z}_\alpha \vc{p}_\beta + \mathcal{Z}_\beta \vc{p}_\alpha)\cdot\frac{\vc{R}}{R^3}
+\frac{\vc{p}_\alpha\cdot\vc{p}_\beta}{R^3}\\\label{Eq2b}
&-3\frac{(\vc{p}_\alpha\cdot\vc{R})(\vc{R}\cdot\vc{p}_\beta)}{R^5}
+ \widetilde{\Phi}_{\alpha\beta}(\vc{R},\zeta),
\end{align}

\begin{align}\nonumber
\widetilde{\Phi}_{\alpha\beta}(\vc{R},\zeta) =& K_\alpha U_\beta(\vc{R},\zeta)
+ K_\beta U_\alpha(\vc{R},\zeta)\\\label{Eq2c}
&+ W_{\alpha\beta}(\vc{R},\zeta) + G_{\alpha\beta}(\vc{R},\zeta).
\end{align}
The first term in Eq. \eqref{Eq2b} describes the long-ranged
ion-ion, the second the dipole-ion
and the third and fourth term the dipole-dipole interaction.
$\mathcal{Z}_\alpha$ and $\mathcal{Z}_\beta$ are the variable charges of the
ions in case the CF's are excited. The latter reduce to the ionic charges
for rigid ions.
$K_\alpha$ and $K_\beta$ are the charges of the ion cores. The remaining
term in Eq. \eqref{Eq2b} given in \eqref{Eq2c} represents the short-ranged interactions.
These short-ranged contributions to the pair potentials are expressed by the following
integrals:

\begin{align}\label{Eq2x}
U_\alpha(\vc{R},\zeta) &= - \int d^3r \rho_\alpha(\vc{r},\zeta)
\left( \frac{1}{|\vc{r}-\vc{R}|} - \frac{1}{R} - \frac{\vc{r}\cdot\vc{R}}{R^3} \right),
\\\nonumber
W_{\alpha\beta}(\vc{R},\zeta) &= \int d^3r \int d^3r'
\bigl[\rho_\alpha(\vc{r},\zeta) \rho_\beta(\vc{r}',\zeta) \times\\\label{Eq2y}
&\times\left(\frac{1}{|\vc{r}-\vc{r}'-\vc{R}|}-\frac{1}{R}-\frac{(\vc{r}+\vc{r}')\cdot\vc{R}}{R^3} \right) \bigr],
\\\nonumber
G_{\alpha\beta}(\vc{R},\zeta) &= \int  d^3r \bigl[\rho_{\alpha\beta}(\vc{r},\zeta)\epsilon(\rho_{\alpha\beta}(\vc{r},\zeta))\\
&-\rho_{\alpha}(\vc{r},\zeta)\epsilon(\rho_{\alpha}(\vc{r},\zeta))\\\nonumber
&-\rho_{\beta}(\vc{r}-\vc{R},\zeta)\epsilon(\rho_{\beta}(\vc{r}-\vc{R},\zeta))\bigr],\label{Eq2z}
\end{align}
with

\begin{equation}\label{Eq2f}
\rho_{\alpha\beta}(\vc{r},\zeta) = \rho_\alpha(\vc{r},\zeta)+\rho_\beta(\vc{r}-\vc{R},\zeta).
\end{equation}

$K_\alpha U_\beta(\vc{R},\zeta)$ yields the short-ranged contribution of the interaction between the
core $\alpha$ and the density $\rho_\beta$ according to Eq. \eqref{Eq1}.
$W_{\alpha\beta}(\vc{R},\zeta)$ represents the short-ranged Coulomb contribution
of the interaction of the density $\rho_\alpha$ with the density $\rho_\beta$.
$G_{\alpha\beta}(\vc{R},\zeta)$ is the sum of the kinetic one-particle- and
the exchange-correlation (XC) contribution of the interaction between the two ions \cite{Ref10}.
The short-ranged part of the potentials and the various coupling coefficients are calculated
numerically for a set of disctances $R$ between the ions. The corresponding results are than
described by an analytical function of the form:

\begin{equation}\label{Eq2d}
f(R) = \pm \text{exp}\left(\alpha+\beta R+\frac{\gamma}{R}\right).
\end{equation}
$\alpha$, $\beta$ and $\gamma$ in Eq. \eqref{Eq2d} are fit parameters.

From the adiabatic condition

\begin{equation}\label{Eq3}
\frac{\partial E(R,\zeta)}{\partial \zeta} = 0
\end{equation}
an expression for the atomic force constants, and accordingly the
dynamical matrix in harmonic approximation can be derived:

\begin{align}\label{Eq4}\nonumber
t_{ij}^{\alpha\beta}(\vc{q}) &=
\left[t_{ij}^{\alpha\beta}(\vc{q})\right]_\text{RIM}\\ &-
\frac{1}{\sqrt{M_\alpha M_\beta}} \sum_{\kappa,\kappa'}
\left[B^{\kappa\alpha}_i(\vc{q}) \right]^{*} \left[C^{-1}(\vc{q})
\right]_{\kappa\kappa'} B^{\kappa'\beta}_j(\vc{q}).
\end{align}

The first term on the right hand side denotes the contribution from
the RIM. $M_\alpha$, $M_\beta$ are the masses of the ions and $\vc{q}$
is a wave vector from the first Brillouin zone (BZ).

The quantities $\vc{B}(\vc{q})$ and $C(\vc{q})$ in Eq. \eqref{Eq4}
represent the Fourier transforms of the electronic coupling
coefficients as calculated from the energy in Eq. \eqref{Eq2}, or the
pair potentials in Eqs. \eqref{Eq2b}-\eqref{Eq2f}, respectivly.

\begin{align}\label{Eq5}
\vc{B}_{\kappa\beta}^{\vc{a}\vc{b}} &= \frac{\partial^2
E(R,\zeta)}{\partial \zeta_\kappa^\vc{a} \partial R_\beta^\vc{b}},
\\\label{Eq6} C_{\kappa\kappa'}^{\vc{a}\vc{b}} &= \frac{\partial^2
E(R,\zeta)}{\partial \zeta_\kappa^\vc{a} \partial
\zeta_{\kappa'}^\vc{b}}.
\end{align}

$\kappa$ denotes the EDF (CF and DF in the
present model, see Eq. \eqref{Eq1}) in an
elementary cell.  The $\vc{B}$ coefficients describe the coupling
between the EDF and the displaced ions (bare electron-phonon
coupling), and the coefficients $C$ determine the interaction between
the EDF.
The phonon frequencies $\omega_\sigma(\vc{q})$ and the
corresponding eigenvectors $\vc{e}^\alpha(\vc{q}\sigma)$ of the modes
$(\vc{q}\sigma)$ are obtained from the secular equation for the
dynamical matrix in Eq. \eqref{Eq4}, i.e.

\begin{equation}\label{Eq7}
\sum_{\beta,j} t_{ij}^{\alpha\beta}(\vc{q})e_j^\beta(\vc{q}) =
\omega^2(\vc{q}) e_i^\alpha(\vc{q}).
\end{equation}

The Eqs. \eqref{Eq4}-\eqref{Eq7} are generally valid and, in
particular, are independent of the specific model for the
decomposition of the perturbed density in Eq. \eqref{Eq1} and the
pair approximation Eq. \eqref{Eq2} for the energy.
The lenghty details of the calculation of the coupling coefficients
$\vc{B}$ and $C$ can not be reviewed in this paper. They are presented
in Ref. \onlinecite{Ref10}. In this context we remark that the
coupling matrix $C_{\kappa\kappa'}(\vc{q})$ of the EDF-EDF
interaction, whose inverse appears in Eq. \eqref{Eq4} for the
dynamical matrix, can be written in matrix notation as

\begin{equation}\label{Eq8}
C = \Pi^{-1} + \widetilde{V}.
\end{equation}

$\Pi^{-1}$ contains the kinetic part to the interaction $C$ and
$\widetilde{V}$ the Hartree and exchange-correlation
contribution. $C^{-1}$ needed for the dynamical matrix and the EPI is
closely related to the (linear) density response function (matrix) and
to the inverse dielectric function (matrix) $\varepsilon^{-1}$,
respectively.

Only very few attempts have been made to calculate the phonon
dispersion and the EPI of the HTSC's using the linear response method
in form of density functional perturbation theory (DFPT) within LDA
\cite{Ref15,Ref16,Ref17}. These calculations correspond to calculating
$\Pi$ and $\widetilde{V}$ in DFT-LDA and for
the \textit{metallic} state only.
On the other hand, in our microscopic modeling DFT-LDA-SIC
calculations are performed for the various densities in Eq. \eqref{Eq1}
in order to obtain the coupling coefficients $\vc{B}$ and
$\widetilde{V}$. Including SIC is particularly important for localized
orbitals like Cu3d in the HTSC's. SIC as a correction for a single
particle term is not a correlation effect, which per definition cannot
be described in a single particle theory, but SIC is important
for contracting in particular the localized Cu3d orbitals.
The latter are correlated orbitals and so we use sometimes in
this paper the phrase ''correlated'' Cu3d states or orbitals,
respectively.
Our theoretical results for the phonon dispersion, which
compare well with the experiments \cite{Ref01,Ref03}, demonstrate that
the approximative calculation of the coupling coefficients in our
approach is sufficient, even for the localized Cu3d states, see also
section \ref{SecThree}. Written in matrix notation we get for the
density response matrix the relation

\begin{equation}\label{Eq9}
C^{-1} = \Pi(1+\widetilde{V}\Pi)^{-1} \equiv \Pi \varepsilon^{-1},
\hspace{.7cm} \varepsilon = 1 + \widetilde{V}\Pi.
\end{equation}

The CF-CF submatrix of the matrix $\Pi$ can approximatively be
calculated for the metallic (but not for the undoped and underdoped)
state of the HTSC's from a TBA of a single particle electronic
bandstructure. In this case the electronic polarizability $\Pi$ in
tight-binding representation reads:

\begin{align}\nonumber
\Pi_{\kappa\kappa'}&(\vc{q},\omega=0) = -\frac{2}{N}\sum_{n,n',\vc{k}}
\frac{f_{n'}(\vc{k}+\vc{q})
-f_{n}(\vc{k})}{E_{n'}(\vc{k}+\vc{q})-E_{n}(\vc{k})}
\times \\\label{Eq10} &\times \left[C_{\kappa n}^{*}(\vc{k})C_{\kappa
n'}(\vc{k}+\vc{q}) \right] \left[C_{\kappa' n}^{*}(\vc{k})C_{\kappa'
n'}(\vc{k}+\vc{q}) \right]^{*}.
\end{align}
$f$, $E$ and $C$ in Eq. \eqref{Eq10} are the occupation numbers, the
single-particle energies and the expansion coefficientes of the
Bloch-functions in terms of tight-binding functions.

The self-consistent change of an EDF on an ion induced by a phonon
mode $(\vc{q} \sigma)$ with frequency $\omega_\sigma(\vc{q})$ and
eigenvector $\vc{e}^\alpha(\vc{q}\sigma)$ can be derived in the form

\begin{equation}\label{Eq11}
\delta\zeta_\kappa^\vc{a}(\vc{q}\sigma) = \left[-\sum_\alpha
\vc{X}^{\kappa\alpha}(\vc{q})\vc{u}_\alpha(\vc{q}\sigma)\right]
e^{i\vc{q}\vc{R}_\kappa^\vc{a}}
\equiv \delta\zeta_\kappa(\vc{q}\sigma)e^{i\vc{q}\vc{R}^\vc{a}},
\end{equation}
with the displacement of the ions

\begin{equation}\label{Eq12}
\vc{u}_\alpha^{\vc{a}}(\vc{q}\sigma) =
\left(\frac{\hbar}{2M_\alpha\omega_\sigma(\vc{q})}
\right)^{1/2}\vc{e}^\alpha(\vc{q}\sigma)e^{i\vc{q}\vc{R}^\vc{a}}
\equiv \vc{u}_\alpha(\vc{q}\sigma)e^{i\vc{q}\vc{R}^\vc{a}}.
\end{equation}

The self-consistent response per unit displacement of the EDF in
Eq. \eqref{Eq11} is calculated in linear response theory as:

\begin{equation}\label{Eq13}
\vc{X}(\vc{q}) = \Pi(\vc{q})\varepsilon^{-1}(\vc{q})\vc{B}(\vc{q}) =
C^{-1}(\vc{q})\vc{B}(\vc{q}).
\end{equation}

A measure of the strength of the EPI for a certain phonon mode
$(\vc{q}\sigma)$ is provided by the change of the self-consistent
potential in the crystal felt by an electron at same space point
$\vc{r}$ in this mode, i.e. $\delta
V_\text{eff}(\vc{r},\vc{q}\sigma)$.  Averaging this quantity with the
corresponding density form factor $\rho_\kappa(\vc{r}-\vc{R}_\kappa^\vc{a})$
at the EDF located at $\vc{R}_\kappa^\vc{a}$, we obtain

\begin{equation}\label{Eq14}
\delta V_\kappa^\vc{a}(\vc{q}\sigma) = \int dV
\rho_\kappa(\vc{r}-\vc{R}_\kappa^\vc{a}) \delta
V_\text{eff}(\vc{r},\vc{q}\sigma).
\end{equation}

This gives a measure for the strength of the EPI in the mode
$(\vc{q}\sigma)$ mediated by the EDF considered. For an expression
of $\delta V_\kappa^\vc{a}(\vc{q}\sigma)$ in terms of the coupling
coefficients in Eqs. \eqref{Eq5}, \eqref{Eq6}, see Ref.
\onlinecite{Ref01}. From our calculations for LaCuO large values
for $\delta V_\kappa^\vc{a}(\vc{q}\sigma)$ are found, in
particular for the phonon anomalies and even larger for the
non-adiabatic $c$-axis phonons in the metallic phase, mixing with
the plasmon.

The generalization for the quantity $\Pi$ in Eqs. \eqref{Eq8}, \eqref{Eq9}
needed for the kinetic part of the charge response in the
non-adiabatic regime, where dynamical screening effects must be
considered, can be achieved by adding $(\hbar\omega+i \eta)$ to the
differences of the single-particle energies in the denominator of the
expression for $\Pi$ in Eq. \eqref{Eq10}.  Other possible non-adiabatic
contributions to $C$ related to dynamical exchange-correlation effects
and the phonons themselves are beyond the scope of the present model.
The coupled-mode frequencies of the phonons and the plasmons must be
determined self-consistently from the secular equation \eqref{Eq4} for
the dynamical matrix which now contains the frequency $\omega$
implicitly via $\Pi$ in the response function $C^{-1}$. Analogously,
the dependence on the frequency is transferred to the quantity
$\vc{X}$ in Eq. \eqref{Eq13} and thus to $\delta\zeta_\kappa$ and
$\delta V_\kappa$ in Eqs. \eqref{Eq11} and \eqref{Eq14},
respectively. Such a non-adiabatic approach is necessary for a
description of the interlayer phonons and the charge-response within a
small region around the $c$-axis \cite{Ref01,Ref04}.

\section{Results and Discussion}\label{SecThree}
\subsection{A modeling of the electronic charge response
and corresponding phonon calculations}

\begin{figure*}
\includegraphics[]{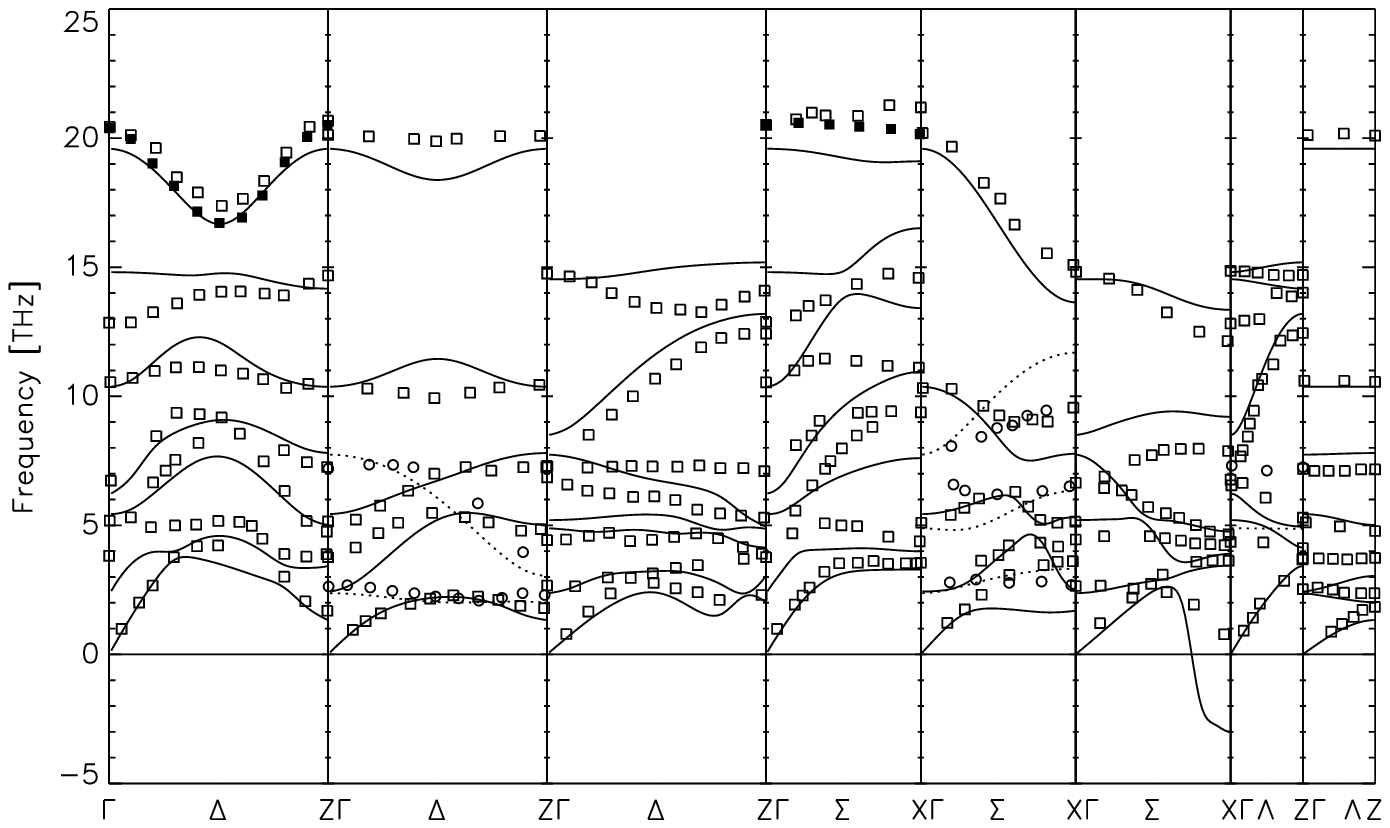}
\caption{Calculated phonon dispersion of LaCuO in
the tetragonal structure in the main symmetry
directions $\Delta \sim (1,0,0)$, $\Sigma \sim (1,1,0)$ and $\Lambda
\sim (0,0,1)$ in the model for the optimally doped state.  The various
symbols representing the experimental results indicate different
irreducible representations (ID's). Open symbols are for
La$_{1.9}$Sr$_{0.1}$CuO$_{4}$ \cite{Ref20} and solid symbols for
La$_{1.85}$Sr$_{0.15}$CuO$_{4}$ \cite{Ref33}. The arrangement of the
panels from left to right according to the different ID's is as
follows:
\newline
  $|$ $\Delta_1$ $|$ $\Delta_2$($\cdot\cdot\cdot$, $\circ$),
  $\Delta_4$($-$, $\square$) $|$ $\Delta_3$ $|$ $\Sigma_1$ $|$
  $\Sigma_2$($\cdot\cdot\cdot$, $\circ$), $\Sigma_4$($-$, $\square$)
  $|$ $\Sigma_3$ $|$ $\Lambda_1$($-$, $\square$),
  $\Lambda_2$($\cdot\cdot\cdot$, $\circ$) $|$ $\Lambda_3$ $|$
  }\label{Fig01}
\end{figure*}

An important question remains as to how to discriminate between the
charge response of the metallic- and the insulating state of the
HTSC's. The latter cannot be obtained within the LDA and a realistic
quantitative description to calculate the irreducible polarization
part $\Pi_{\kappa\kappa'}(\vc{q})$ for the HTSC's is not
available. However, a general criterion follows from the different
analytic behaviour of the polarizability in the long-wavelength limit
$(\vc{q}\rightarrow\vc{0})$ in both phases \cite{Ref01,Ref05}. In the
\textit{metallic phase} the electronic partial
density of states (PDOS) at the
Fermi level $Z_\kappa(\varepsilon_\text{F})$ is related to the
polarizability matrix for $(\vc{q}\rightarrow\vc{0})$ according to

\begin{equation}\label{Eq15}
\sum_{\kappa'} \Pi_{\kappa\kappa'}(\vc{q}\rightarrow\vc{0}) =
Z_\kappa(\varepsilon_\text{F}),
\end{equation}
and the total density of states at energy $\varepsilon$ is given by

\begin{equation}\label{Eq16}
Z(\varepsilon) = \sum_{\kappa} Z_\kappa(\varepsilon).
\end{equation}

On the other hand, for the \textit{insulating} state we obtain the sum rules

\begin{equation}\label{Eq17}
\sum_{\kappa'} \Pi_{\kappa\kappa'}(\vc{q}\rightarrow\vc{0}) =
\mathcal{O}(q)
\end{equation}
and

\begin{equation}\label{Eq18}
\sum_{\kappa\kappa'} \Pi_{\kappa\kappa'}(\vc{q}\rightarrow\vc{0}) =
\mathcal{O}(q^2).
\end{equation}

The sum $\sum_{\kappa\kappa'}
\Pi_{\kappa\kappa'}(\vc{q}\rightarrow\vc{0})$ is equal to $\rho^2K$ with
$\rho$ the average density and $K$ the compressibility of the electronic
system. The latter provides a measure of the
gap in the electronic spectrum because $K$ vanishes as a function of the
chemical potential in the gap region. Equations \eqref{Eq15}, \eqref{Eq16}
and \eqref{Eq17}, \eqref{Eq18}, respectively, can be considered as an
\textit{orbital resolved} closed form to describe the metal-insulator
transition in terms of the polarizability or the compressibility of
the electronic system, respectively. This
is a primary tool to characterize
the corresponding ground state.  Such orbital selective sum rules are
particularly useful in the proximity to a Mott insulating phase where
different from the band insulator the internal degrees of freedom,
orbital (and spin), approximatively survive. In order to fulfil the
sum rules above, in case of the insulating state, in contrast to the
metallic state, \textit{off-diagonal} elements
of the polarizability matrix
describing \textit{nonlocal} polarization processes
necessarily must occur and
interfere in order to correlate the charge response in such a way that
Eqs. \eqref{Eq17}, \eqref{Eq18} are satisfied.

Physically this is related to the fact that in an insulator a
perturbation, e. g. a change of the electron-ion potential, is
only incompletely screened. Consequently, the selfconsistent
change of the potential at the orbital degree of freedom $\kappa'$
nonlocally contributes to the CF's of the orbital
$\kappa\neq\kappa'$ in the unit cell. This is quite different from
the metallic case where the diagonal elements of
$\Pi_{\kappa\kappa'}$ dominate and off-diagonal elements can be
neglected in most cases. As mentioned in the Introduction, the
model for the polarizability of the metallic state is based on LDA
calculations which, however, have to be modified to correct for
the overestimation of the $c$-axis coupling in LDA-based
investigations \cite{Ref03}. So, a model taking into account Cu3d,
4s, 4p, and O$_\text{xy}$2p orbitals in the CuO
plane results \cite{Ref03}. In
contrast to the Cu4s orbital the effect of the Cu4p orbital on the
phonon dynamics is only weak, so we have neglected the latter in
the present calculations and enhanced the contribution of $\Pi(Cu4s)$
slightly. Besides the CF's also anisotropic DF's
are taken into account.

The resulting calculated phonon dispersion is shown in Fig.
\ref{Fig01}. We get an overall agreement with the experimental
dispersion curves. A few branches show larger deviations between
experiment and theory. However, in particular the important phonon
anomalies (high-frequency oxygen bond-stretching modes), which are
a major theme in the recent literature and the present paper, too,
are well described by our theory, see Fig.\ref{Fig04} below.
This also holds true for the steep $\Lambda_1$ branch which results
from the large anisotropy of the real material and cannot be
described with pure LDA based calculations leading to an
overestimation of the coupling along the $c$-axis,
see Ref. \onlinecite{Ref03}  Freezing-in
the unstable $\Sigma_3$ tilt mode at the $X$ point correctly
indicates the experimentally observed structural phase transition
from the high-temperature tetragonal (HTT) to the low-temperature
orthorhombic (LTO) structure.  The actual transition, of course,
cannot be studied in the harmonic approximation because the
anharmonic contribution to the energy must
be considered \cite{Ref18}. However,
from our calculations in Ref. \onlinecite{Ref01} we find that the
soft tilt mode found in HTT is stabilized in LTO
in harmonic approximation and thus the
structural transition is essentially driven by the long-ranged
ionic forces in the material.

\begin{figure*}
\includegraphics[]{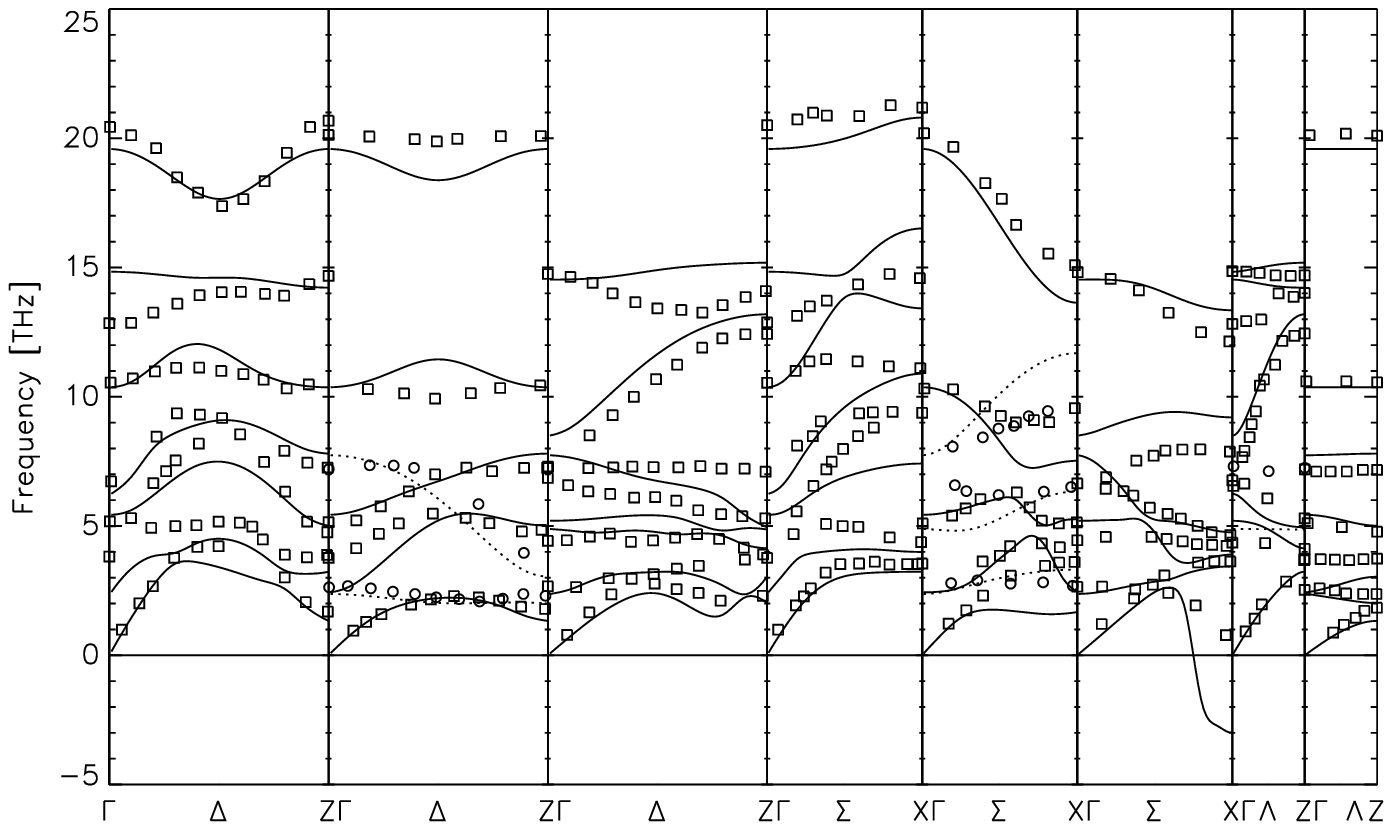}
\caption{Same as Fig. \ref{Fig01} with the calculated results from the
model for the underdoped state.}\label{Fig02}
\end{figure*}

The non-empirical potential-induced breathing (PIB) model \cite{Ref19}
like our model reflects the strong tendencies to ionic forces in the
HTSC's. This model also correctly predicts, that the tetragonal structure
is unstable to the $X$-point tilt mode. However, it can not be applied
to the metallic phase. Moreover, other unstable phonon branches are found
in this model. The width of the phonon spectrum is
considerably overestimated. For example, the PIB result for the planar oxygen
breathing mode at the $X$-point in the insulating phase is about 34 THz in
comparsion to the experiment ($\sim$ 21.8 THz) and to our calculated result of
21.6 THz, see Fig. \ref{Fig04}. The PIB results are comparable to our
calculations performed with the ab-initio rigid ion model using nominal
ionic charges \cite{Ref05}, neglecting the covalence correction for the ionic
charges and the pair potential. Taking the latter effects into account, as
mentioned in Section \ref{SecTwo}, the correct width of the spectrum is
obtained and, with the exception of the tilt mode at $X$, all unstable
phonon branches disappear \cite{Ref12}. So, we obtain a suitable reference
system for the HTSC which allows for the investigation
of the additional nonlocal,
non-rigid screening effects in terms of CF's and DF's.

Empirical shell model calculations for the lattice dynamics of several cuprates
have been performed in Ref. \onlinecite{Ref20}. The fitting to the experimental
results is quite satisfactorily emphasizing again the importance of the
ionic forces. However, in the fitting the harmonic $X$-point tilt-mode
was assumed to be stable, which is at least questionable, because
of the low temperature orthorhombic structure.
For a description of the metallic phase free-carrier screening
of the homogeneous electron gas is additional assumed, but
the softening of the phonon anomalies can not be accounted for
by this type of screening. On the other
hand, our calculations, see Fig. \ref{Fig01},
lead to very good results for the phonon anomalies.
Accordingly, we can relate the anomalous softening to the
characteristic ionic, inhomogenous
electronic structure of the HTSC, leading to strong
nonlocal EPI in terms of localized CF's in the outer shells of
the ions controlled by the coupling coefficient $C$ in Eq. \eqref{Eq8}

In the model for the insulating phase in order to simulate
stronger correlation effects as in the metal Cu4s orbitals are
neglected and only the CF's of the Cu3d orbitals at the Cu and O2p
orbitals at the O$_\text{xy}$ are considered as on site CF's and
the polarizability is adjusted so as to obey the sum rules from
Eq. \eqref{Eq17}. The calculated results for the phonon dispersion
can be found in Ref. \onlinecite{Ref03}. Similarily as for the
metallic state a good agreement with the experiments is obtained.
In particular, a significant reduction of the phonon anomalies
as compared to the optimally doped material is well reproduced
and attributed to the incompressibility of the electronic state.
Including the screening contribution via anisotropic DF's,
the longitudinal and transverse optical modes of $A_{2u}$ and
$E_u$ symmetry are well described.

An ab initio calculation of the polarizability $\Pi$ for the
HTSC's, which besides strong electron correlations also should
include the strong nonlocal EPI found in our calculations in the
self-energy of the electrons, seems not possible in the near
future. In particular this holds for the insulating and underdoped
state. However, modeling $\Pi$ according to the the sum rules
\eqref{Eq15} - \eqref{Eq18} guarantees that the constraints on the
charge response in the different phases are fulfilled and,
moreover, allows for the possibility to theoretically design
condensed matter states with properties between that of a
conventional metal and an insulator, respectively. This seems
necessary to describe the physics in the underdoped cuprates where
a strange metallic state does appear. Our modeling of the
correlation phenomenon and the delocalization-localization
transition from the metallic state of the HTSC's via the
underdoped state to the undoped (insulating) state has been
achieved with help of the sum rules by consecutive orbital
selective compressibility-incompressibility transitions, see Ref.
\onlinecite{Ref02}, and for a detailed review Ref.
\onlinecite{Ref01}. Starting from the metallic side, representing
in our model the doping range from optimally doped to overdoped,
all the admitted orbitals, i.e. Cu3d, 4s and O2p, are taken to be
compressible, metallic consistent with an emerging quasi-particle
description and a well defined Fermi surface. The corresponding
sum rules appropriate for such a situation are given by Eqs.
\eqref{Eq15}, \eqref{Eq16} with \textit{all} the
$Z_\kappa(\varepsilon_\text{F})$ different from zero. Our
calculation of the orbital occupation numbers \cite{Ref12} from a
TBA of the first principles electronic (metallic) bandstructure of
LaCuO \cite{Ref21} indicates that besides the Cu3d in particular
the more delocalized Cu4s state must be included as an additional
EDF at the Cu site. This, as already mentioned, is different from
our modeling of the insulating - and underdoped state where we
have neglected a contribution from the Cu4s orbitals and only used
the localized Cu3d - and the O2p orbitals as EDF's in order to
simulate stronger correlation effects in these states.

Approaching the delocalization-localization transition from the
metallic region when p-doping is decreased first the Cu3d
component of the wavefunction is admitted in our model to become
incompressible, insulator-like in the underdoped regime by
assuming that the corresponding single particle density of states
$Z_\kappa(\varepsilon)$ is suppressed at the Fermi level for the
correlated localized Cu3d orbitals, because of the cost in energy
from hopping of the charge carrier to the Cu sites, but not so for
the more delocalized O2p orbitals at the O$_\text{xy}$-sublattices
in the CuO plane where the holes are predominantly injected to in
p-type cuprates. So, we have an orbital selective compressible,
metallic charge response for the O2p orbitals with a renormalized
PDOS $\widetilde{Z}_\kappa(\varepsilon_\text{F})$ and the holes
tend to accumulate in the region of suppressed antiferromagnetism.
Altogether, we have a loss in the density of states (pseudogap,
correlation gap) at the Fermi level. Thus, the quasiparticle
picture ($Z_\kappa(\varepsilon_\text{F}\neq 0)$ for \textit{all}
$\kappa$) consistent with our modeling of the metallic state
for optimally to overdoping no
longer holds in the underdoped state. The related
compartmentalization of phase space can be expected to be the
reason for the strange metallic behaviour observed in the
underdoped cuprates with an ill defined Fermi surface. The
corresponding electronic state for this model (pseudogap model,
PGM) expressing the dichotomy between a localized and delocalized
component has been defined by the following sum rule \cite{Ref02},

\begin{equation}
\sum_{\kappa'} \Pi_{\kappa\kappa'}(\vc{q}\rightarrow\vc{0}) =
\begin{cases}
\mathcal{O}(q) \hspace{.7cm} &\text{Cu3d}\\
\widetilde{Z}_\kappa(\varepsilon_\text{F})&\text{O2p}.
\end{cases}
\end{equation}

We realize from this equation, that nonlocal polarization processes
are necessary to depress the density of states
at $\varepsilon_\text{F}$ via the Cu3d component
and correspondingly the low energy single
particle excitations.
Finally, near half filling in addition to Cu3d, the O2p component of
the electronic state has been modeled as incompressible too, and the
charge-transfer insulator is formed according to the sum rules
\eqref{Eq17}, \eqref{Eq18} for both the Cu3d and O2p orbitals.

The phonon dispersion curves for our model of the underdoped state
with an incompressible insulator-like, charge response of the Cu3d
orbital at the Cu sublattice and a compressible, metallic charge
response of the O2p orbitals at the O$_\text{xy}$-sublattice in
the CuO plane is displayed in Fig. \ref{Fig02} and compared with
experimental inelastic neutron results for underdoped LaCuO.  The
agreement of the calculations, which also include anisotropic DF's
besides the CF's of Cu3d- and O2p type, is good and of similar
quality as the calculated results for optimally doped state in
Fig. \ref{Fig01} and the insulating state in Ref.
\onlinecite{Ref03}. Comparing the calculations in Figs.
\ref{Fig01}, \ref{Fig02} essential differences are only found for
the phonon anomalies (highest $\Delta_1$ and $\Sigma_1$-branch,
respectively) which are less pronounced in the underdoped state,
because the partial ''\textit{correlation gap}'' introduced by
both, the incompressible Cu3d orbitals and the missing of the
compressible Cu4s orbitals leads according to Eq. \eqref{Eq11} to
reduced insulator-like CF's at the Cu ion. Compared with the
insulating state, however, the anomalies are increased in the
underdoped phase in particular for the $\Delta_1$-branch. A more
detailed discussion of the phonon anomalies is given below.
Another difference of the theoretical results as compared with
those of the insulating state given in Ref. \onlinecite{Ref03} is
of \textit{qualitative} nature and is related to the axial
polarized $\Lambda_1$-branches. In the modeling of the underdoped
state we have partially incompressible regions, however, the total
compressibility is never zero in this state, because a real space
organization of the low lying charge excitations is allowed and a
metallic (gapless) charge transport is
enabled via the hole-doped O2p
orbitals at the oxygen network in the CuO plane. As a consequence
the $A_{2u}$ mode splittings are closed due to metallic screening,
in particular the very large ''ferroelectric'' mode splitting,
observed and calculated in the insulating state \cite{Ref03,Ref04}
and the $\Lambda_1$-branch with the steep dispersion
characteristic for metallic behaviour, being absent in the
insulating state, reappears as is visible from Fig. \ref{Fig02}.
Altogether, we conclude that our modeling of the insulating,
underdoped and optimally doped state of LaCuO is consistent
with the measured phonon dispersion.

\subsection{Qualitative discussion of the modeling
of the electronic state in the cuprates}

We now discuss some qualitative aspects of our modeling of the
electronic state of the cuprates suggested by the phonon calculations.
In the model for the underdoped
state a metallic mobility behaviour is mediated via the O2p component
of the wavefunction localized at the oxygen network of the CuO plane
where the holes are injected too, but blocked along the Cu-O links
because of the large on-site repulsion of the Cu3d component. Thus, by
this partial ordering we have hole rich and hole poor regions of the
conducting carriers on a microscopic scale in a translational
invariant way ultimately driven by competing energy
contributions. Note, that the various inhomogeneous selforganized
charge and spin ordered configurations discussed in the litertature
\cite{Ref22,Ref23,Ref24} formally can be considered as certain
''defect structures'' (e.g. stripes) of our model if the spin-degrees
of freedom are explicitly considered.  In case such structures are
experimentally observed they express the tendency to restore an
insulating Mott state in certain extended (incompressible) regions of
the material.

From an energetic point of view the ground state should simultaneously
minimize the zero point kinetic energy of the doped charge carriers
and the exchange energy where both compete which each other. Because
of the incompressible localized Cu3d states
the holes are expelled from the
antiferromagnetic regions by Coulomb correlation. Thus, an independent
particle picture where an electron feels a time-averaged local density
of the other electrons with a corresponding local potential (like
e.g. in LDA) no longer holds.
In this way the exchange energy is optimized (no
spins need to be switched because of the
restricted motion of the holes) while on the other
hand, there is a gain in kinetic energy in the hole rich regions by
delocalization of the electronic structure (hopping of the holes) via
the compressible, metallic O2p component of the wavefunction. At the
same time the exchange energy can be neglected in this region of
configuration space if the delocalization effect is strong
enough. Spin degrees of freedom and their correlations are not
explicitly included in our modeling but implicitly support the latter
because residual antiferromagnetic correlations can lead to an
additional lowering of the kinetic energy by virtually hopping to the
Cu3d states. This correlates the charge transport in the regions of
excess charge to the remaining antiferromagnetic spin order in the
hole poor regions at the Cu sublattice. So, we have a picture
that can be regarded as resulting from a kind of separation
of spin and charge, being mutually relevant in different
regions of the CuO plane. Low energy charge excitations occur
in the hole rich regions where spin effects are depressed but are
blocked in the hole poor regions, where the spin degrees of freedom
are most important at low energy and high-energy (virtual) CF's
induce an antiferromagnetic coupling on the Cu sublattice.

Then, in our modeling of the metallic state at higher doping the
Cu3d and the additionally allowed delocalized Cu4s component
become compressible, metallic. The incompressibility of Cu3d can
be thought to be destroyed by an increased population of the Cu4s
orbital. In Ref. \onlinecite{Ref25} it has been shown that the Cu4s
admixture to the electronic states of the HTSC is relevant and
due to a gain in correlation energy. The dominant charge
transfer resulting from the atomic correlations is from the localized
Cu3d to the more delocalized Cu4s orbitals. The calculations
in Ref. \onlinecite{Ref25} are based on an ab-initio method
(local ansatz method) that adds correlations as corrections
to a single-particle self-consistent ground state obtained from
a Hartree-Fock calculation. An increased hybridization of Cu4s
with the $pd\sigma$ orbitals can be expected to further relaxing
the kinetic energy when doping is increased.
As a consequence, the blocking of the
metallic charge response at the Cu sites in the underdoped state
is lifted and metallic behaviour is now also possible via the CuO
links.  Altogether, this seems to be consistent with an emerging
quasiparticle picture, a progressively weakening of the
antiferromagnetic fluctuations and the development of a well
defined Fermi surface as described by the sum rules from Eqs.
\eqref{Eq15}, \eqref{Eq16} in our approach. The growing importance
of the Cu4s component in the wavefunction leading to enhanced
delocalized CF's at the Cu sites in the screening process is also
mapped to a corresponding renormalization of the phonon anomalies
as compared to the situation where the latter are neglected, see
Ref. \onlinecite{Ref26} and the discussion below. So, from our
calculation additionally to Cu3d and O2p an increasing
contribution of the delocalized Cu4s component to the metallic
state of the cuprates is important to understand the charge
response and the corresponding phonon softening
in the real material at least at higher doping levels.

The importance of the Cu4s orbital for a realistic description of
the electronic structure of the HTSC's has also been pointed out
in Ref. \onlinecite{Ref27} and Ref. \onlinecite{Ref28}.
In the latter work, the hopping range has
been identified as an essential material-dependent parameter and
the intralayer hopping beyond nearest neighbours as well as the
interlayer hopping proceeds via the Cu4s orbital. It ist further
concluded that materials with higher $T_{C,\text{max}}$ have
larger hopping ranges and in materials with highest
$T_{C,\text{max}}$ the axial orbital, essentially a hybrid between
Cu4s, Cu3d$_{3\text{z}^2-1}$ and apical oxygen 2p$_\text{z}$, is
almost pure Cu4s. Moreover, the importance of Cu4s for an accurate
description of the partial charge distributions in the HTSC's is
pointed out applying the ''local ansatz'' ab initio method, which
allows to add correlations as corrections to a Hartree-Fock ground
state\cite{Ref25} or density functional
theory \cite{Ref29}, respectively.

Now we address the question concerning the possible nature of a
superconducting state of the HTSC's that would be consistent with
our modeling of the normal state. The suppression of the local
CF's in our model for the underdoped state related to the
incompressible Cu3d orbitals due to the large on-site repulsion
competes with the superconductivity because of the enhanced phase
fluctuations according to the number-phase uncertainty
relationship $\Delta N \Delta \phi \geq 1$. In such a phase
disordered superconductor pairing without long-range phase
coherence is possible, i.e. there is an intermediate temperature
range $(T > T_C)$ where electron pairs exist which have not
condensed.  Such a precursor pair formation scenario is for
example discussed in context with a kinetic energy driven pairing
mechanism related to certain self organized local charge
inhomogenities in form of stripes \cite{Ref22,Ref23} where pairing
originates from strong repulsive interactions. Pair formation is
thought to be realized in form of spin singlet pairs resulting in
a spin gap by the ''spin gap proximity'' effect \cite{Ref24}
because it seems favourable that, under appropriate circumstances,
the pairs can spread out somewhat into the antiferromagnetic
neighbourhood of a stripe. However, up to now there is no
experiment that proves the pairing is either kinetic energy driven
or due to stripes. Note in this context, that the stripe scenario
has been questioned by recent experiments which find that the
observed nanoscale disorder in the HTSC's is due to dopant
disorder \cite{Ref30}. Moreover, if the possibility of precursor
pairs of some kind is theoretically explored, the strong nonlocal
EPI found in our calculations expressed by the corresponding large
changes of the self-consistent potential an electron feels
according to Eq. \eqref{Eq14}, in particular for the generic
phonon anomalies \cite{Ref31} and even larger for the
non-adiabatic $c$-axis phonons \cite{Ref01,Ref04}, should be taken
into account in the electronic self-energy and the polarizability
besides the short-ranged electron-electron repulsion that leads to
spin- and charge correlations in a many body treatment of a
preformed pair state and the resulting gaps in the normal as well
in the superconducting state. For a recent review of the important
role of the EPI for the normal state properties and the pairing in
the HTSC's we also refer to Ref. \onlinecite{Ref32}. Altogether,
in the normal state one can expect two types of gaps in our
modeling for the underdoped HTSC's the ''correlation gap'' and
consistent with that modeling possibly a ''precursor pairing gap''
which are both linked by the short-ranged correlations introduced
through the localized, incompressible Cu3d orbitals. While the
former gap should correspond to the high-energy pseudogap seen in
ARPES, the latter might be related to the low-energy gap found by
the same experimental technique \cite{Ref06}.

Different from the underdoped state in our modeling of the
optimally doped metallic state \textit{all} the orbitals become
compressible, metallic with enhanced CF's particularly in the Cu4s
orbitals. So, we have a qualitative change to an electronic ground
state with a well defined Fermi surface. In such a state a Fermi
surface instability with a simultaneous evolution of pairs of
quasi particles would be consistent with the normal state and can
lead to a BCS-type ordering. With regard to predominantly a phase
ordering transition in the underdoped state optimal doping with a
maximal $T_C$ can be thought of as defining a gradual crossover to
an overdoped state with predominantly a BCS-type pairing
transition. Arguments in favour of such a crossover scenario also
have been presented in Ref. \onlinecite{Ref22,Ref23}.

According to the modeling of the charge response supported by our
phonon calculations there are higher energy CF's related to the
localized Cu3d orbital and low energy CF's of the more delocalized O2p
- and especially the Cu4s orbitals and a critical mixing of both via a
population of Cu4s and a corresponding change of the populations of
Cu3d and O2p and the related hybridization obviously is required to
achieve $T_{C,\text{max}}$ in the optimally doped material. The
localization-delocalization transition in the normal state expressed
as an incompressibility-compressibility transition of the Cu3d states
corresponds to a phase ordering-pairing transition in the
superconducting state.

\subsection{Doping dependent anomalous phonon branches
- Nonlocal dynamic charge inhomogenities}
Next, we report new calculations of the phonon dispersion for the
overdoped state of LaCuO which can be compared with recent
experimental results for the phonon anomalies \cite{Ref33}. An
important message from these calculations is a further growing of the
importance of the delocalized Cu4s component in the electronic wave
function which seems consistent with an improved Fermi liquid
quasi-particle picture, an increased coupling along the $c$-axis and
the dying out of the antiferromagnetic correlations, and an EPI which
becomes more local.

\begin{figure*}
\includegraphics[]{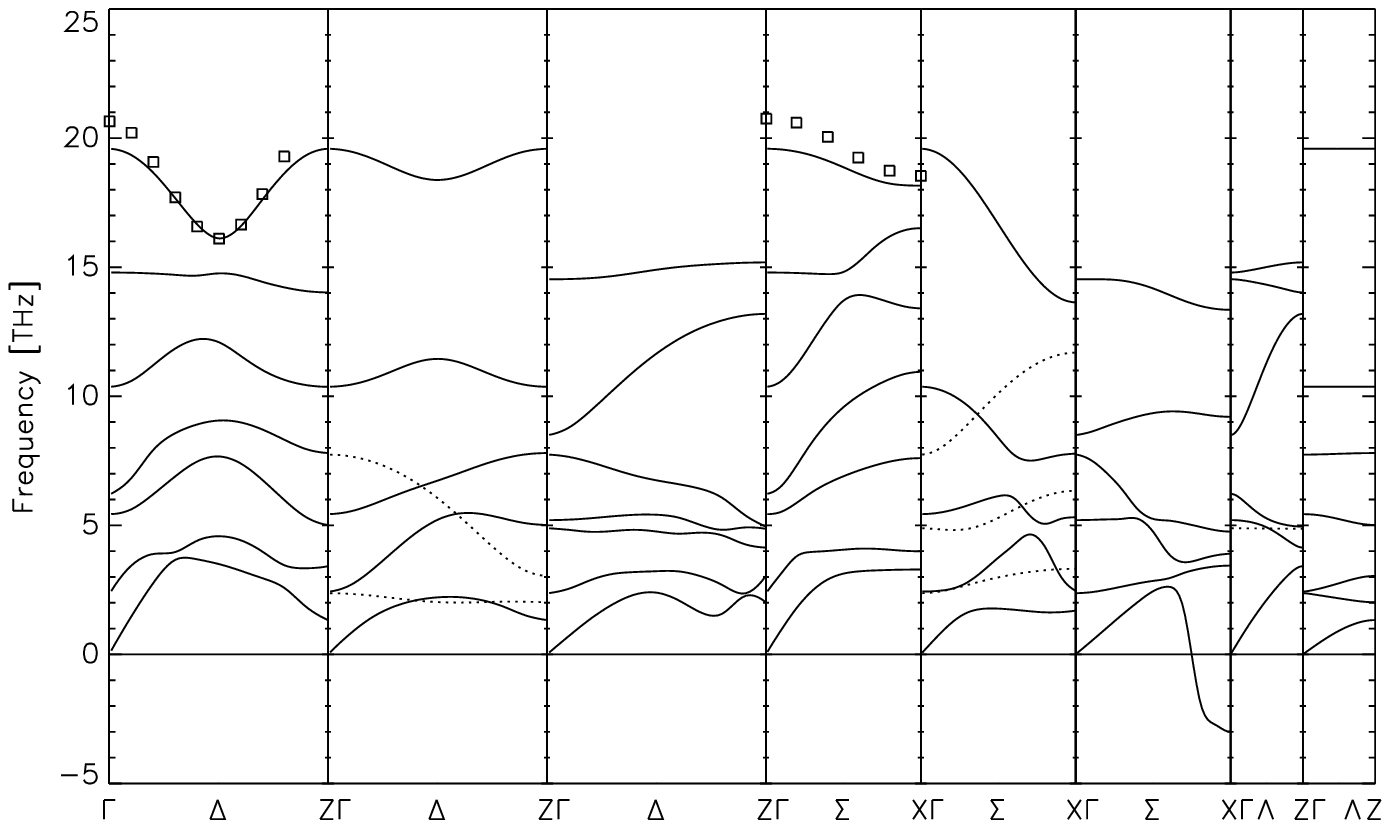}
\caption{Calculated phonon dispersion of LaCuO in the main symmetry
directions as in Fig. \ref{Fig01} within the model for the overdoped
state. The symbols indicate the experimental results for
La$_{1.7}$Sr$_{0.3}$CuO$_4$ of the branches with the phonon anomalies
\cite{Ref33}.}\label{Fig03}
\end{figure*}

In earlier calculations \cite{Ref01,Ref26} we already have recognized
the inclusion of Cu4s into the orbital basis as an important ''tuning
parameter'' for the delocalization of the electronic structure of the
HTSC's and the strength of the phonon anomalies. A growing occupation
of the Cu4s orbital and a corresponding growth of the Cu(4s)-element
of the polarizability matrix leads to an increased softening of the
anomalous phonon modes. So, enhancing $\Pi(\text{Cu4s})$ as compared
with its value in the optimally doped state
leading to the results shown in
Fig. \ref{Fig01}, gives very good agreement with the values measured
for the anomalies in the overdoped state \cite{Ref33} and, moreover,
provides a prediction for the complete phonon dispersion not measured
so far, see Fig. \ref{Fig03}.

\begin{figure*}%
      \centering
      \includegraphics[]{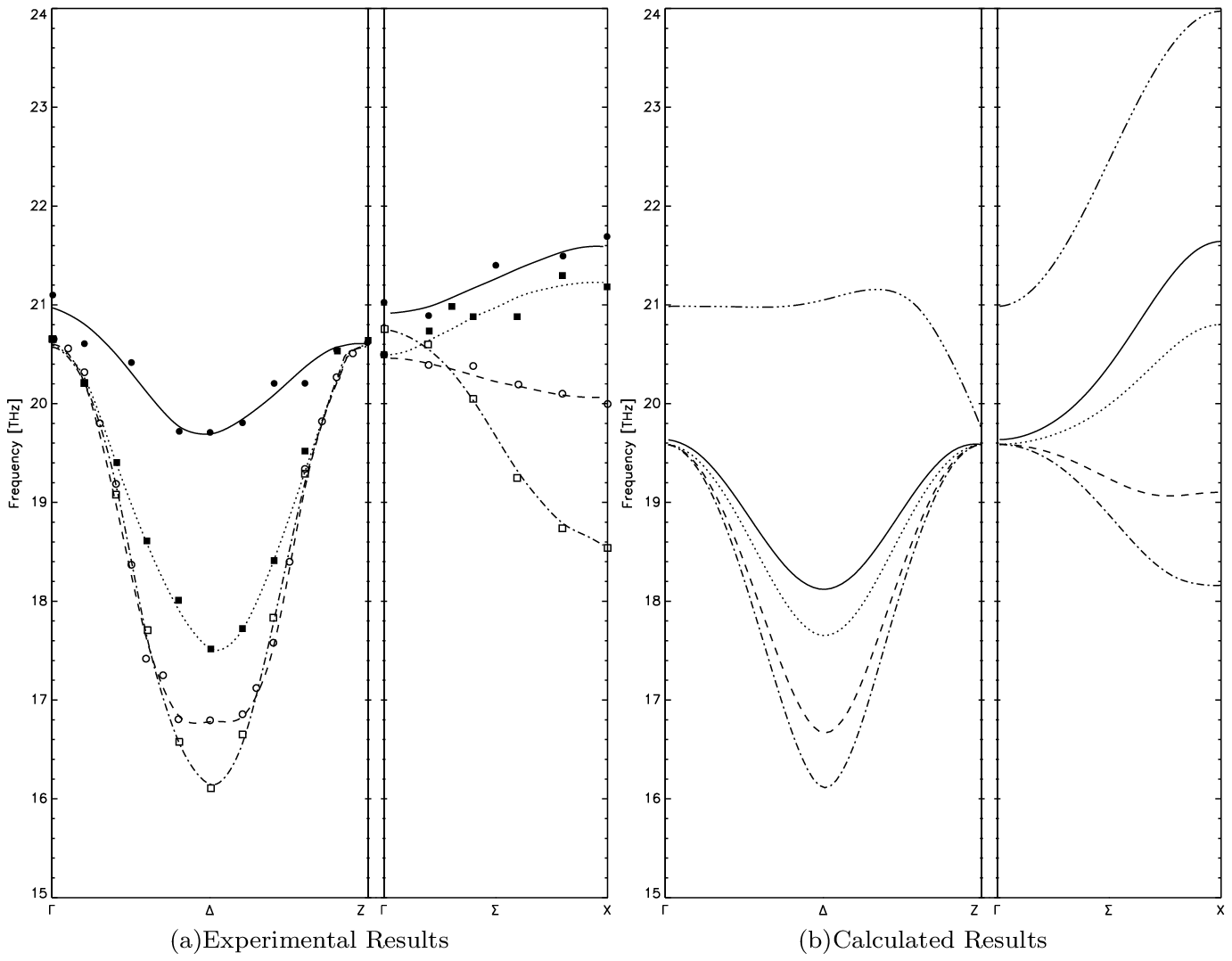}%
  \caption{(a) Experimental results for the highest $\Delta_1$ and
  $\Sigma_1$ branch of La$_{2-x}$Sr$_x$CuO$_4$ for various doping
  levels \cite{Ref33}: $\bullet x=0.00$, $\blacksquare x=0.10$, $\circ
  x=0.15$, $\square x = 0.30$.  The lines are a guide to the eye. (b)
  Calculated results of the phonon branches shown in (a) as obtained
  within our model approach for the electronic state of the HTSC's:
  $-$ insulating state, $\cdots$ underdoped state, - - optimally doped
  state, $-\cdot-$ overdoped state. For comparison, the results for
  the ab initio RIM are shown ($-\cdots$) in order to demonstrate the
  large influence of the nonlocal EPI effects in form of CF's on the
  dispersion.}\label{Fig04}%
\end{figure*}%

\begin{figure}
\includegraphics[width=\linewidth]{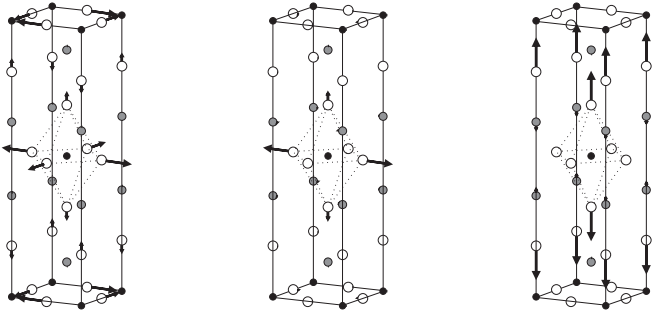}
\caption{Displacement patterns of the high-frequency oxygen
bond-stretching modes of LaCuO. Left: planar breathing mode
(O$_\text{B}^X$), middle: half-breathing mode ($\Delta_1/2$), right:
apex-oxygen breathing mode (O$_\text{z}^Z$).}\label{Fig05}
\end{figure}

In Fig. \ref{Fig04} we show a comparison of our calculated results for
the anomalous phonon branches, $\Delta_1$ and $\Sigma_1$, in LaCuO
with the corresponding experimental dispersion \cite{Ref33}.  The
strong doping dependence of the experimental phonon frequences and
thus of the EPI is clearly visible and very well accounted for by our
modeling of the electronic state of the HTSC's discussed above. We
have also included the phonon dispersion as obtained for the RIM
representing approximatively the ionic local part of the charge
response and the EPI in order to emphasize the large contribution due
to the nonlocal EPI and screening in terms of the CF's. The
displacement patterns of the phonon anomalies, i e. the planar oxygen
breathing mode at the $X$ point, O$^X_\text{B}$, and the oxygen
half-breathing mode, $\Delta_1/2$, are displayed in
Fig. \ref{Fig05}. For symmetry reasons there are only CF's at the
silent Cu ion allowed in O$^X_\text{B}$ and not on the moving
O$_\text{xy}$ ions while in $\Delta_1/2$ in addition to the CF's at
the Cu ion also CF's can be excited at the silent oxygens. This leads
to an additional source of softening in case of $\Delta_1/2$ as
compared to O$^X_\text{B}$ and the larger renormalization of the
frequencies found in our calculations
and the experiments \cite{Ref33,Ref34}. As already
mentioned the enhanced softening in the overdoped doped state can be
attributed to the growing Cu4s component in the electronic state and
the related CF's at low energy. Moreover, the local character of the
EPI is increased being dominant in a normal metal and superconductor,
with delocalized electrons.

A word should be said concerning the importance of a correct
representation of the on-site Coulomb repulsion, $U_d$, related to the
localized Cu3d states and their effect on the nonlocal EPI
by suppressing low energy CF's.
The onsite-parameters, like $U_d$ for the Cu3d orbitals, are defined
in our modeling using Eq. \eqref{Eq6} and
taking the EDF $\kappa$, $\kappa'$
to be localized at the same ion $\alpha$ in the crystal. An explicit
expression for the calculation of these parameters is given by
the following equations\cite{Ref10}.

\begin{align}\nonumber
U_{\kappa\kappa'} =& \int dV \int dV' \frac{\rho_\kappa(\vc{r})\rho_{\kappa'}(\vc{r}')}{|\vc{r}-\vc{r}'|}\\\label{Eq25}
&+ \int dV \rho_\kappa(\vc{r}) \rho_{\kappa'}(\vc{r}) w \left[\rho_\alpha^0(\vc{r}) \right],
\end{align}

with

\begin{align}
w(\rho) = \frac{d\mu(\rho)}{d \rho}\text{,\hphantom{AAAAA}}
\mu(\rho) = \frac{d\left[\rho\epsilon(\rho)\right]}{d\rho}.
\end{align}

The relevant form factors
$\rho_\kappa$, $\rho_{\kappa'}$ in Eq. \eqref{Eq25} are defined
by the $\rho^\text{CF}_\lambda$ from Eq. \eqref{Eq1} and
$\rho_\alpha^0$ is the unperturbed density of the ion $\alpha$
calculated in LDA-SIC as noted in Section \ref{SecTwo}.
The calculated value for $U_d$ in units of $e^2/a_\text{B}$ is
1.0050. The corresponding value for the more delocalized O2p and
Cu4s states is 0.6708 and 0.3563, respectively.
The large value for the Cu3d orbitals tends to supress CF's at
the Cu site according to Eqs. \eqref{Eq8}, \eqref{Eq11}
and \eqref{Eq13}, while the latter are strongly enhanced by the much
smaller value for the Cu4s orbital. This again points to the
important role of a Cu4s admixture in the electronic state and
for the anomalous phonons in the cuprates.

\begin{figure}
\includegraphics[width=0.99\linewidth]{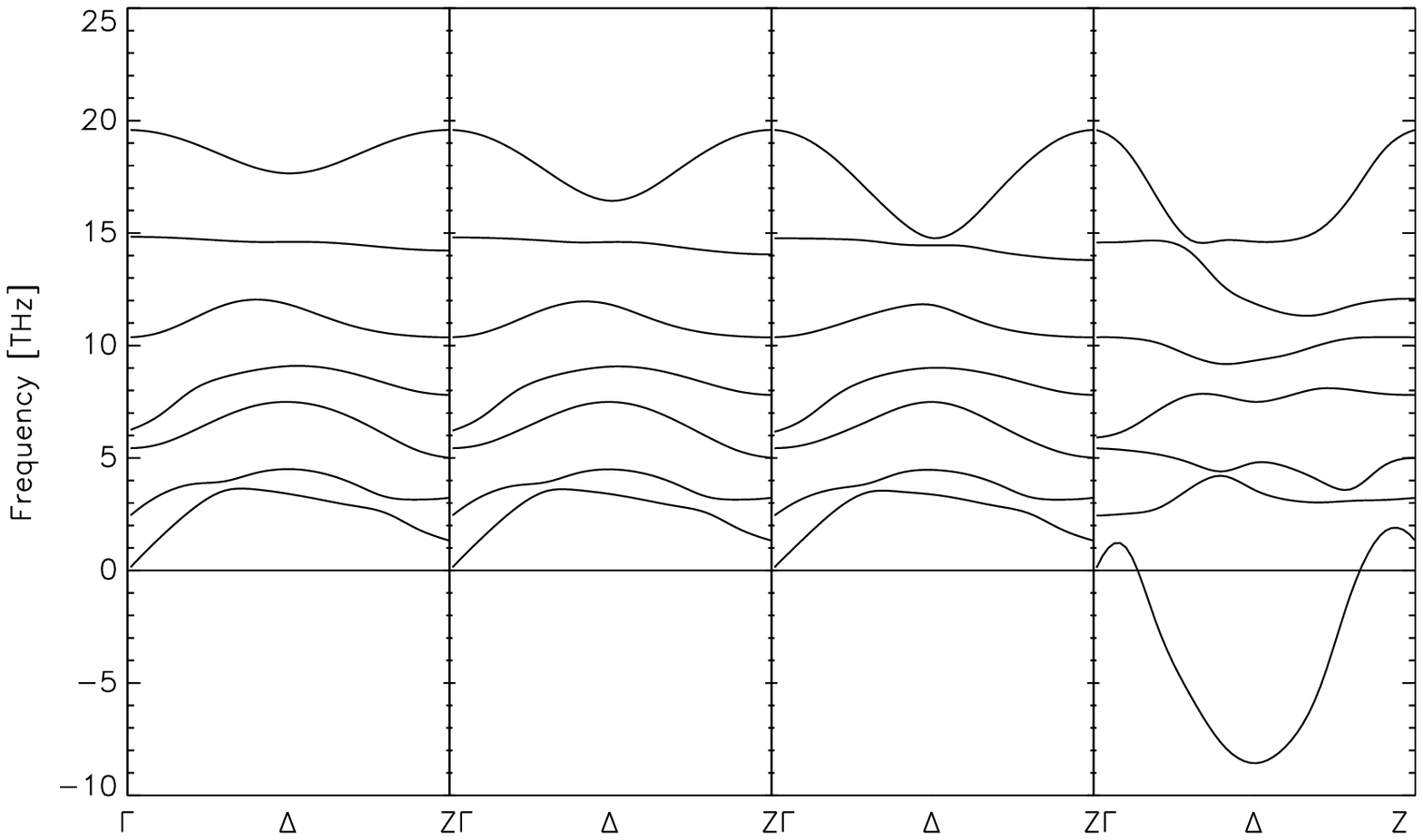}
\caption{Demonstration of the selective sensibility of the
anomalous highest $\Delta_1$ branch with respect to the on-site
Coulomb repulsion $U_d$ of the localized Cu3d orbitals. Only
$\Delta_1$ branches coupling to CF's are shown for the model of
underdoped state of LaCuO. Using the calculated value for $U_d$
(DFT-LDA-SIC, left panel) good agreement with the experimental
data (Fig. \ref{Fig02}) is obtained. In the remaining panels from
left to right we arbitrarily have decreased $U_d$ by the factor
$\frac{2}{3}$, $\frac{1}{2}$ and $\frac{1}{3}$, respectively. For
$\frac{1}{3}U_d$ the frequency of the $\Delta_1/2$ anomaly
($\approx$4.8 THz) is decreased towards an instability. In a
corresponding calculation for the $\Sigma_1$ branches
O$_\text{B}^X$ even becomes unstable for
$\frac{1}{3}U_d$.}\label{Fig06}
\end{figure}

A characteristic feature of the phonon
anomalies is their selective sensitivity to $U_d$. A decrease of $U_d$
enhances the CF's at the Cu ion considerably and leads in parallel to
a softening of the phonon anomalies while the other phonon branches
coupling to CF's are virtually not affected; see Fig. \ref{Fig06} for
the $\Delta_1/2$ anomaly, where we have decreased $U_d$ from its
calculated value which by itself leads to the good quantitative
agreement with the experiments in Fig. \ref{Fig02}. A similar
calculation for the $\Sigma_1$ branches even leads to an unstable
planar breathing mode O$^X_\text{B}$ being the highest mode in the
insulating and underdoped state (other phonon branches coupling to
CF's are only weakly modified). So, we conclude that our approximative
calculation of $U_d$ within DFT-LDA-SIC is sufficient to describe the
correlation effect related to $U_d$ on the phonon dynamics.

Local charge transfer from Cu3d into Cu4s orbitals due to atomic
correlations has been shown in Ref. \onlinecite{Ref25} to lead to
a strong reduction (screening) of $U_d$
for effectiv model Hamiltonian
calculations. The influence on the phonon anomalies in LaCuO of
such a screening effect for $U_d$ already has been investigated in
Ref. \onlinecite{Ref05}. In this work a model with a reduced set
of EDF is used and CF's are only allowed for
the Cu3d and the O$_\text{xy}$2p states of the ions in the CuO
plane. The contribution of more delocalized states, like Cu4s or
Cu4p, is simulated in this model by
assuming a renormalizing (screening) of the
on-site interaction $U_d$. Within such a procedure, the anomalous
softening for $\Delta_1/2$ and O$^X_\text{B}$ now seen in the
experiments has been predicted and related to the changes of the
effective repulsive short-range Coulomb
interaction at the Cu ion as an
important physical parameter of the cuprates.

From an ab-initio point of view the definite value for an effectiv
$U_d$ parameter which should be used in models with a reduced
set of electronic degrees of freedom, like the Hubbard models,
which is widely used in the cuprates is by no means straightforward.
Quite generally, a screening of the Coulomb interaction can be
discussed within a \textit{renormalization scheme} of the electronic
density response as proposed in \cite{Ref11,Ref35,Ref36}, leading
for the inverse dielectric matrix $\varepsilon^{-1}$ to the
following \textit{identity}

\begin{equation}\label{Eq20}
\varepsilon^{-1} = \varepsilon^{-1}_\text{r} \cdot \bar{\varepsilon}^{-1},
\end{equation}

with

\begin{equation}\label{Eq21}
\varepsilon^{-1}_\text{r} = \left(1 + \bar{v} \Delta \right)^{-1},
\hspace{.4cm}\bar{\varepsilon}^{-1} = \left(1 + v \bar{\Pi} \right)^{-1},
\hspace{.4cm}\bar{v} = \bar{\varepsilon}^{-1}v.
\end{equation}

This \textit{general} factorization of the screening follows from
an additive decomposition of the total polarizability $\Pi$ into
a suitable chosen renormalizing contribution ($\bar{\Pi}$) and a
remaining part , $\Delta$, that contains all the polarization
processes not included in $\bar{\Pi}$, i.e.

\begin{equation}\label{Eq22}
\Pi = \bar{\Pi} + \Delta.
\end{equation}

Applying Eq. \eqref{Eq20} to the bare Coulomb interaction, $v$,
we obtain the screened interaction

\begin{equation}\label{Eq23}
V_\text{SC} = \varepsilon^{-1} v = \left(1 + \bar{v}\Delta \right)^{-1}\bar{v}.
\end{equation}

In \eqref{Eq22} we can for example identify $\Delta$ with that part
of $\Pi$ which contains only 3d-3d transitions and $\bar{\Pi}$ with
the remaining part of the total polarizability, i.e. by the other
(renormalizing) electronic orbitals considered. Using this specific
realization of the decompostion \eqref{Eq22}, we can extract from
Eq. \eqref{Eq23} that the interaction between the 3d electrons
is provided by the interaction $\bar{v}$ defined by the screening
processes of the other orbitals taking into account in $\Pi$.

In general the interaction between the 3d electrons, $\bar{v}$,
becomes frequency-dependent (retarded) via the frequency-dependence
of the polarizability $\bar{\Pi}$, which could be calculated e.g.
at the RPA level, like in our treatment of the non-adiabatic regime.
Summarizing, the interaction in a model reduced to the 3d degrees
of freedom should be \textit{frequency-dependent} in contrast to
the \textit{static} $U_\text{d}$-parameter assumed intuitively in
Hubbard-like Hamiltonian approaches. From our considerations a
static on-site interaction $U_\text{d}$ should be identified with
the partially renormalized interaction $\bar{v}$ in the low-energy limit,
i.e. $U_\text{d} = \bar{v}(\omega\rightarrow 0)$.
In this context we refer to recent work where it is found that a static
value of $U_\text{d}$ may not be the most appropriate one to use
in effective model calculations\cite{Ref43}.

In summary, from our calculations we find that the phonon anomalies
and the strong nonlocal EPI can be understood in terms of excited CF's
in the outer shells of the ions, which are controlled by the kinetic
contribution $\Pi^{-1}$ to the electronic coupling coefficient $C$ in
Eq. \eqref{Eq8} together with the contribution $\widetilde{V}$,
i.e. the Hartree- and exchange-correlation part of $C$. The interplay
between strong correlation effects related to the localized Cu3d
orbitals with their tendency to suppress CF's with low energy via
$\widetilde{V}$ and the contribution of the more extended orbitals O2p
and Cu4s with a smaller (calculated) on-site repulsion which tends to
enhance CF's is an essential feature for an understanding of the
charge response and its influence on the phonons in a quantitative
manner.

We have made explicit the charge response by calculating the
phonon-induced charge density redistribution \cite{Ref01,Ref02,Ref26}
for the phonon anomalies. According to these investigations the
nonlocal EPI generates dynamic charge ordering via CF's in form of
localized stripes of alternating sign in the CuO plane leading to an
effective reduction of the energy visible in the anomalous
softening. For the $\Delta_1/2$ mode the charge stripes point along
the $x$- or $y$-axis, respectively, and for O$_\text{B}^X$ along the
diagonals in the CuO plane. These dynamical charge inhomogeneities
would not be present in systems with local EPI, for example in a high
density homogeneous electron gas prevailing in conventional metals. On
the other hand, certain charge inhomogeneities observed in the HTSC's
because of the reduced screening will couple nonlocally to the lattice
and additional softening, compared with the system free of
inhomogeneities, of the oxygen-bond-stretching-like modes (OBSM) can
be expected due to the sensitivity of the latter to charge doping as
seen in Fig. \ref{Fig04}. For example, in the presence of a periodic
charge order at a certain wavevector compatible with its period,
anomalous softening and related mode splitting according to the broken
rotational and translational symmetry should occur. Also non-periodic,
local charge inhomogeneities could lead to anomalous local OBSM via
non-local EPI. These theoretical expectations are supported by
recent experiments \cite{Ref37}. Here strong phonon softening of the
OBSM has been observed in HTSC's at doping levels associated with
static stripes at a wave vector corresponding to the charge
order. From the calculated results of the anomalies in
Fig. \ref{Fig04} for the system with unbroken symmetry it can be
expected that the strength of the anomalies in the inhomogeneous case
sensitively should depend on the possibility of the inhomogeneous
structure (e. g. stripe) to be incompressible, insulating or
compressible, metallic, respectively.

It is interesting to note that the electron correlation which
result in an antiferromagnetic spin correlation in the HTSC's
should favour the CF's in the OBSM and enhance screening against
the tendency of a suppression of the CF's by the large on-site
repulsion at the Cu, because if the spins in neighbouring Cu
orbitals were to be parallel, they would not both to be able to
transfer to the neighbouring oxygens or between the Cu ions
themselves. On the other hand, if they are antiparallel they can
do. Thus the phonon induced CF's depend implicitly on the
spin-degrees which may enhance the former and the EPI in an
antiferromagnetic neighbourhood as compared to a paramagnetic or a
ferromagnetic one. Vice versa the nonlocal EPI of CF-type is
expected to generate antiferromagnetic spin-fluctuations via the
phonon-induced transferred charge between the Cu ions. This
discussion around the phonon anomalies provides an example to
illustrate, how coupled lattice-charge- and spin degrees of
freedom could act synergetically for pairing in the cuprates in a
doping dependent way.

Besides $\Delta_1/2$ and O$_\text{B}^X$ essentially polarized in
the CuO plane another OBSM, i.e. the apex-oxygen (O$_\text{z}$)
bond-stretching mode at the $Z$ point, (O$_\text{z}^Z$), polarized
perpendicular to the CuO plane, (Fig. \ref{Fig05}), is of special
interest for the cuprates due to the long-range Coulomb
interaction in these compounds. Here we have the situation
(similar as for the La$_\text{z}^Z$ mode) that the displacement of
the ions (O$_\text{z}$, La) in the ionic layers generates changes
of the potential felt by the electrons in the CuO plane which
themselves are responsible for the superconductivity. Such
nonlocal coupling effects are an expression of the strong
component of the ionic binding along the $c$-axis in the HTSC's,
i.e. these long-range Coulomb coupling effects are very special to
this class of materials, and would not be possible in a convential
metal or superconductor because of local-screening by a
high-density electron gas. From the experiments in LaCuO
\cite{Ref33,Ref34} we get the very remarkable result that O$_\text{z}^Z$
is strongly renormalized when going from the insulating parent
compound ($\approx$17 THz) to the optimally doped material
($\approx$11.5 THz). However, in the optimally doped probe a very
large linewidth of about 4 THz appears that makes it difficult to
localize the exact position of O$_\text{z}^Z$ in the metallic
phase. The massive line broadening has been interpreted within our
non-adiabatic calculations \cite{Ref04} leading to a
phonon-plasmon scenario for modes propagating in a very small
region parallel to the $c$-axis. From the calculations
\cite{Ref01,Ref04} we find a small region nearby the $c$-axis with
a non-adiabatic, insulator-like charge response crossing over to a
coherent adiabatic, metallic response outside this region.
Moreover, an interconnection between the in-plane response and the
size of the non-adiabatic inter-plane response is pointed out. So
the electronic properties in the CuO plane take influence on the
charge response perpendicular to the plane nearby the $c$-axis.

\begin{figure}
\includegraphics[width=\linewidth]{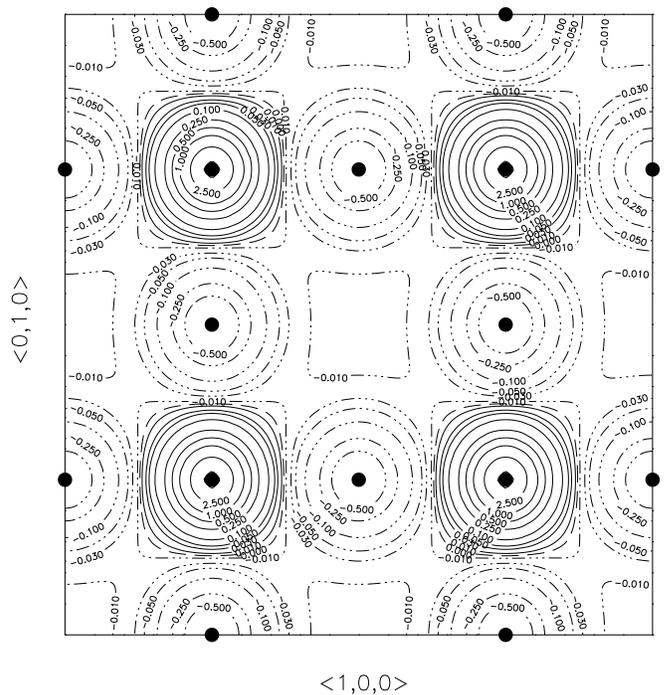}
\caption{Contour plot of the nonlocal part of the phonon-induced
charge density redistribution for the O$_\text{z}^Z$ mode
(Fig. \ref{Fig05}) in the model for the insulating state of LaCuO. The
units are 10$^{-4}e^2/a_\text{B}^3$. Full lines ($-$) mean that
electrons are accumulated in the corresponding region of space and the
lines ($-\cdots$) represent regions where the electrons are pushed
away, resulting in an \textit{intra-layer} charge
transfer.}\label{Fig07}
\end{figure}

\begin{figure}
\includegraphics[width=\linewidth]{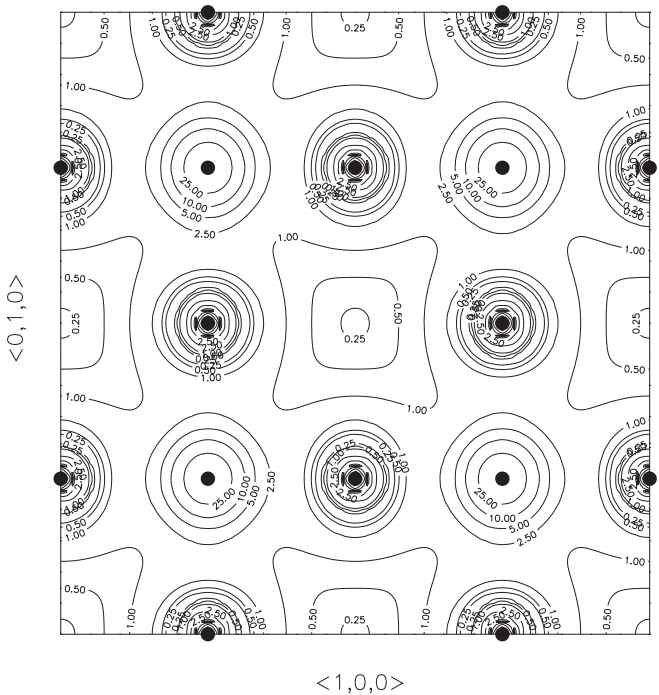}
\caption{Same as Fig. \ref{Fig07} but calculated with the model
representing the optimally doped metallic state, resulting in CF's of
the same sign in a CuO layer and a corresponding \textit{inter-layer}
charge transfer.}\label{Fig08}
\end{figure}

The strong softening in the metallic state outside the small
region of non-adiabatic charge response (for details see Ref.
\onlinecite{Ref01,Ref04}) can physically be understood by
comparing the phonon induced charge rearrangements for
O$_\text{z}^Z$ as calculated in the insulating and metallic state
(Figs. \ref{Fig07}, \ref{Fig08}), respectively. For the
frequencies we find a renormalization of about 3 THz from 17.13
THz in the insulator to 14.17 THz in the metal, compare with the
adiabatic phonon dispersion in Fig. 1 where O$_\text{z}^Z$ is the
second highest frequency of the $\Lambda_1$-branches at the $Z$
point. A large EPI for the O$_\text{z}^Z$ mode in the metallic
state of LaCuO has also been calculated in Ref. \onlinecite{Ref38}
and attributed to a combination of weak screening along the
$c$-axis and nonlocal Madelung-like interactions.
The importance of such long-ranged ionic contributions to coupling
also has been pointed out in Refs. \onlinecite{Ref39,Ref40,Ref44,Ref45}.

As can be extracted from Fig. \ref{Fig05} for O$_\text{z}^Z$ the
apex oxygens move in phase against the CuO layers. So, because of
the weak screening one can expect this vibration to induce CF's in
the CuO planes. These CF's, however, are \textit{qualitatively}
different in the insulating and the metallic state, respectively,
because of the gap for charge excitations in the insulator. This
helps to understand the anomalous softening of the mode during the
insulator-metal transition. As has been shown in Ref.
\onlinecite{Ref05} the CF's $\delta\zeta_\kappa(\vc{q}\sigma)$
from Eq. \eqref{Eq11} are constrained in the insulator with CF's
allowed at the Cu and O$_\text{xy}$ sublattices according to the
following sum rule

\begin{equation}\label{Eq24}
\sum_\kappa \delta\zeta_\kappa(\Lambda\sigma) = 0.
\end{equation}

The sum over $\kappa$ denotes the CF's in the CuO layer.
$\Lambda\sim(0,0,1)$, so Eq. \eqref{Eq24} particularly holds at
the $Z$ point of the Brillouin-zone. In contrast to the constraint
expressed by Eq. \eqref{Eq24} for the insulating state, which
means that local charge neutrality of the cell is maintained under
a perturbation due to O$_\text{z}^Z$, no such a restriction is
present for the metallic state. Unlike for LaCuO in the insulating
state, where only \textit{intra-layer} charge rearrangements
according to the sum rule Eq. \eqref{Eq24} are allowed (Fig.
\ref{Fig07}), in the metallic state O$_\text{z}^Z$ induces CF's at
Cu and O$_\text{xy}$ that have the same sign in the whole CuO
layer (Fig. \ref{Fig08}).  This finally leads to CF's of
alternating sign in consecutive CuO layers, i.e. to an
\textit{inter-plane} charge transfer in the adiabatic regime that
provides an effective screening mechanism for the long-range
Coulomb interactions and generates the anomalous mode softening of
O$_\text{z}^Z$ during the insulator-metal transition.

\begin{figure*}%
\includegraphics[]{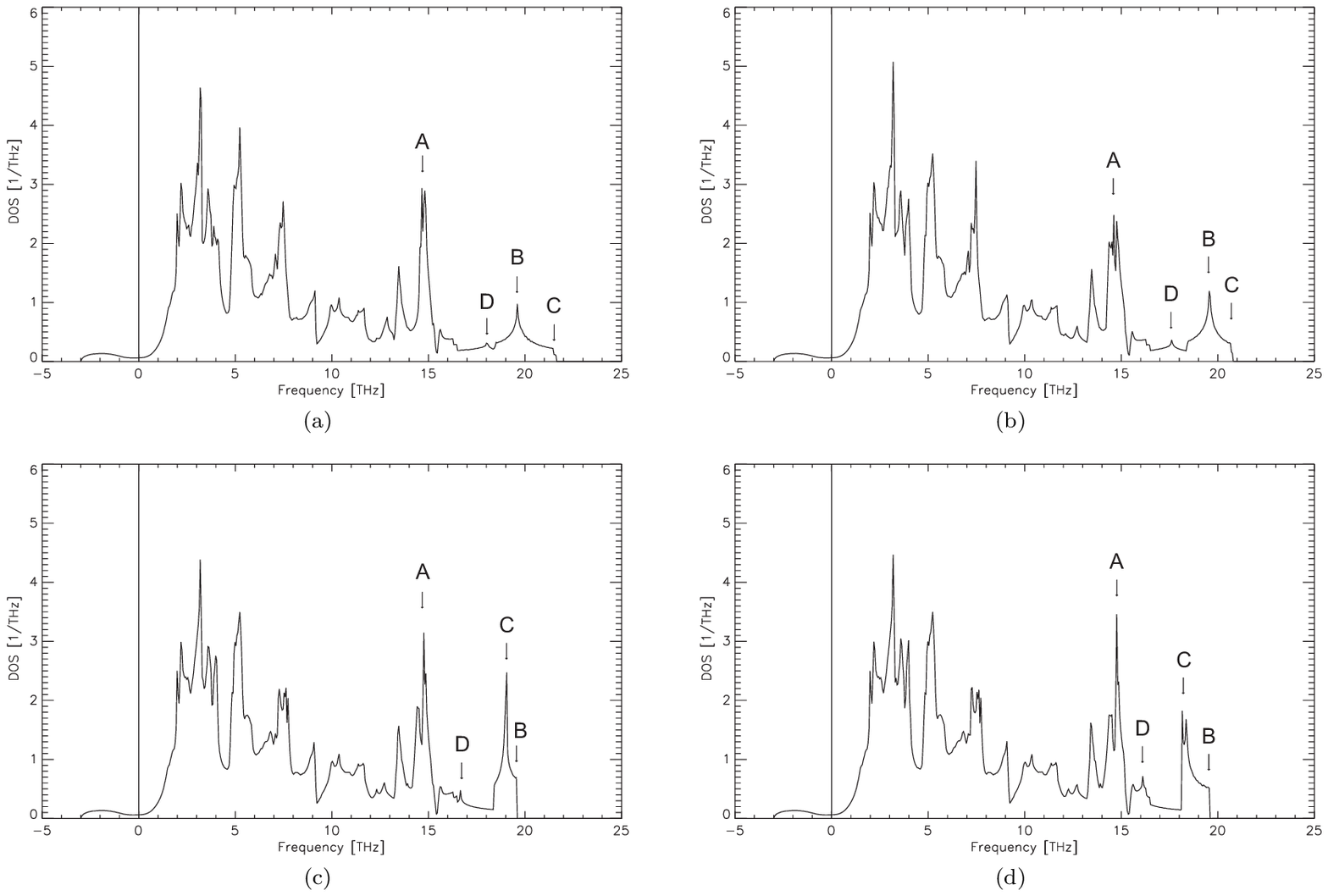}%
\caption{Phonon density of states for LaCuO as calculated with the
model for the insulating state (a), the underdoped state (b), the
optimally doped state (c) and the overdoped state (d),
respectively.}\label{Fig09}%
\end{figure*}%

\subsection{Phonon-density of states}
In the last topic of this paper we present calculations for the
phonon-density of states of LaCuO according to our modeling of the
insulating-, under-, optimally- and overdoped state of the HTSC's.
Experimental studies by inelastic neutron spectroscopy of the
phonon-density of states of LaCuO report unusual shifts for
high-frequency OBSM during the insulator-metal transition
indicating a strong EPI \cite{Ref41}. Similar results are found
for LaCuO in Ref. \onlinecite{Ref42} where besides a shift of the
high-frequency part of the spectrum to lower frequencies the
development of ''new"''oxygen lattice vibrations centered in the
optimally doped probe around 70 meV are observed near the doping
induced insulator-metal transition and attributed to the mixing of
these vibrations with metallic CF's.

\begin{figure*}%
\includegraphics[]{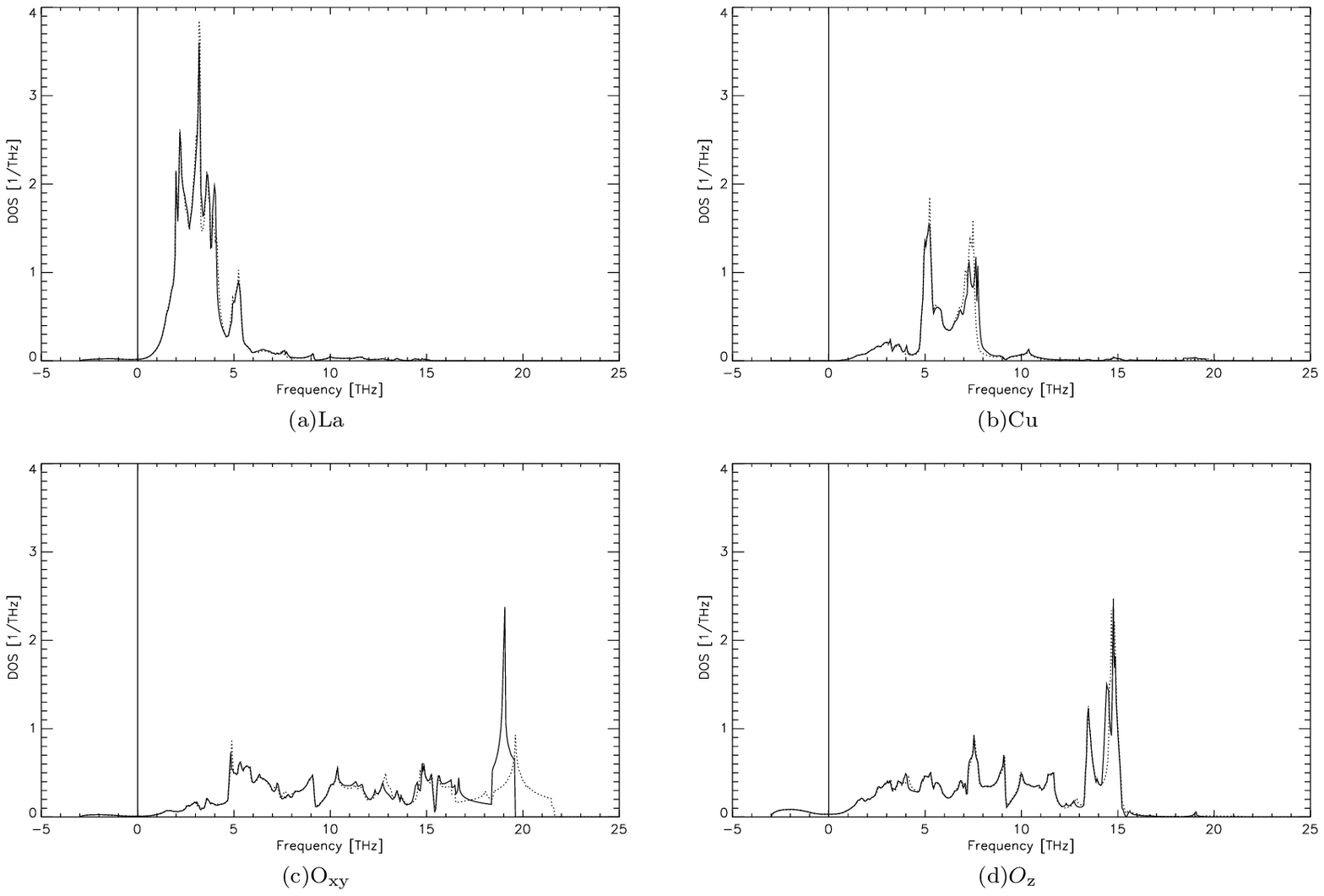}%
\caption{Comparison of the atom resolved phonon density of states of
LaCuO between the insulating state $\cdots$ and the optimally doped
metallic state $-$.}\label{Fig10}%
\end{figure*}%

Our calculations for the phonon-density of states displayed in
Figs. \ref{Fig09}, \ref{Fig10} allow for a detailed study of the
changes of the spectrum during the insulator-metal transition.
With the help of the calculated atom-resolved phonon-density of
states in Fig. \ref{Fig10} and the phonon dispersion the
characteristic modes indicated in Fig. \ref{Fig09} can be
identified. The peak denoted as $A$ is caused by vibrations of the
apex-oxygen, O$_\text{z}$. $B$ is related to the high-frequency
oxygen vibrations, $E_u$, at the $\Gamma$ point polarized in the
CuO plane. $C$ is the planar oxygen breathing mode,
O$_\text{B}^X$, and $D$ results from the oxygen half-breathing
mode ($\Delta_1/2$ mode). In agreement with the experiments we
find globally a shift of the high-frequency part of the
phonon-density of states and also the development of spectral
weight (peak $D$, see also Fig. \ref{Fig10} (c)) passing from the
insulating to the metallic state which we can assign to the
$\Delta_1/2$ anomaly at 69 meV in our calculations for the
optimally doped phase. Thus, the ''new'' oxgen lattice vibrations
found in the experiments consist dominantly of the half-breathing
modes, as supposed in Ref. \onlinecite{Ref42}. An important change
of the spectrum related to the insulator-metal transition can be
extracted from Figs. \ref{Fig09}, \ref{Fig10}, because $B$($E_u$)
and $C$(O$_\text{B}^X$) are interchanged. So we have not just a
global softening but also a very characteristic rearrangement of
the high-frequency part of the spectrum, linked to the development
of the phonon anomalies and the corresponding softening upon
doping.

\section{Summary}\label{SecFour}
Within our microscopic modeling of the insulating-, underdoped-,
optimally- and overdoped electronic state of the HTSC's we have
performed calculations for LaCuO of the complete phonon dispersion
curves, dynamic charge inhomogeneities induced by nonlocal EPI and the
total-and atom-resolved phonon density of states.  Our calculations
emphasize the role of the doping dependence of the generic phonon
anomalies and their selective sensitivity to both localization
(correlation) effects provided by the Cu3d orbitals and delocalization
effects via the Cu4s states.
Thus, these anomalies are suitable probes to mirror the
localization-delocalization transition of the electronic structure of
the cuprates in terms of phonons.
We have found good agreement with the
corresponding measured phonon dispersion and phonon-density of states
and have presented an interpretation of the results.
In order to reach these results, both, long-ranged Coulomb interactions
of ionic origin as well as short-ranged repulsiv interactions together
with a sufficient set of orbital degrees of freedom are neccesary. A
purely ionic model leads to a considerable overestimation of the width
of the phonon spectrum and to unstable branches besides the tilt mode.
Including additionally covalent corrections a suitable starting model
for the HTSC can be found that cures the disadvantages of the pure
ionic description. Using such a model as an unprejudiced \textit{rigid}
reference system the effects of the important nonlocal \textit{non-rigid}
electronic polarization processes in terms of localized CF's and DF's
have been studied. The latter localized screening effects and not
a homogenous electron gas screening allows for a quantitative calculation
of the phonon anomalies and the dispersion. By the present
investigations our earlier studies have been extended to the
underdoped and overdoped state. Predictions of the dispersion in the
latter case have been made.

An important aspect of the calculations,
besides the efforts to explore the phonon dynamics and the EPI in the
HTSC's as strongly correlated systems, where these topics presently
are not well understood, in particular in the insulating- and
underdoped state, is to give additional support to our modeling of the
electronic state. This is achieved in terms
of consecutive orbital selective
incompressibility-compressibility transitions. Such a modeling of the
normal state qualitatively also provides a possible picture of the
superconducting state that points from a phase ordering transition in
the underdoped regime to a BCS-like pairing transition for higher
doping. Optimal $T_C$ should mark a gradual crossover point. The
localization-delocalization crossover in our approach corresponding to
a phase ordering-pair ordering crossover in the superconducting state
is essentially related to the increase of an occupation with doping of
the delocalized Cu4s orbitals progressively mixing into the Cu3d-O2p
dominated electronic state of the underdoped material. Simultaneously
an emerging quasiparticle picture with a well defined Fermi surface is
consistent with our modeling upon doping. Finally, the strong and
material specific nonlocal EPI effects found in the calculations
demonstrate the important role of the phonons for the physics of the
cuprates.

\end{document}